\documentclass[12pt]{article}
\pdfoutput = 1
\usepackage{latexsym}
\usepackage{epsfig}
\usepackage{amsmath}
\usepackage{amssymb}
\usepackage{epsf}
\usepackage{epsfig}
\usepackage{cite}
\usepackage{cancel}
\usepackage{accents}
\usepackage{enumerate}
\newcommand{\lsim}{\
\mathrel{\hbox{\rlap{\hbox{\lower4pt\hbox{$\sim$}}}\hbox{$<$}}}}

\newcommand{\gsim}{\
\mathrel{\hbox{\rlap{\hbox{\lower4pt\hbox{$\sim$}}}\hbox{$>$}}}}
\newcommand{\tev}{\, {\rm TeV}}
\newcommand{\gev}{\, {\rm GeV}}
\newcommand{\mev}{\, {\rm MeV}}

\newcommand{\be}{\begin{equation}}
\newcommand{\ee}{\end{equation}}

\def\Bsmumu{B_s\to\mu^+\mu^-}

\begin{document}
\begin{titlepage}
\vspace*{-0.0truecm}

\begin{flushright}
{FLAVOUR(267104)-ERC-35}\\
Nikhef-2013-007
\end{flushright}

\begin{center}
\boldmath
{\Large{\bf Probing New Physics with the $\Bsmumu$ Time-Dependent Rate }}
\unboldmath
\end{center}

\vspace{0.3truecm}

\begin{center}
{\bf Andrzej~J.~Buras$^{a,b}$, Robert Fleischer$^{c,d}$, \\
    Jennifer Girrbach$^{a,b}$ and Robert Knegjens$^{c}$
 \\[0.4 cm]}

\vspace{0.5truecm}

{\small
$^a$TUM Institute for Advanced Study, Lichtenbergstr. 2a, D-85748 Garching, Germany\\
$^b$Physik Department, Technische Universit\"at M\"unchen,
James-Franck-Stra{\ss}e, \\D-85748 Garching, Germany\\
$^c$ Nikhef, Science Park 105, NL-1098 XG Amsterdam, Netherlands \\
$^d$ Department of Physics and Astronomy, Vrije Universiteit Amsterdam, NL-1081 HV Amsterdam, Netherlands}
\vskip0.71cm

\end{center}

\vspace{1.0cm}
\begin{abstract}
\vspace{0.2cm}\noindent

The $\Bsmumu$ decay plays an outstanding role in tests of the Standard Model and physics beyond it. 
The LHCb collaboration has recently reported the first evidence for this decay at the $3.5\,\sigma$ level, with a 
branching ratio in the ballpark of the Standard Model prediction. Thanks to the recently established sizable decay width difference of the $B_s$ system, another observable, ${\cal A}^{\mu\mu}_{\Delta\Gamma}$, is available, which can be extracted from the time-dependent untagged $\Bsmumu$ rate. If tagging information is available, a CP-violating asymmetry, ${\cal S}_{\mu\mu}$, can also be determined. These two observables exhibit sensitivity to New Physics that is complementary 
to the branching ratio. We define and analyse scenarios in which these quantities allow us to  
discriminate between model-independent effective operators and their CP-violating phases.
In this context we classify a selection of popular New Physics models into the considered scenarios.
Furthermore, we consider specific models with tree-level FCNCs mediated 
by a heavy neutral gauge boson, pseudoscalar or scalar,
finding striking differences in the predictions of these scenarios for the observables considered and the correlations among them.
We update the Standard Model prediction for the time-integrated branching ratio taking 
the subtle decay width difference effects into account. We find $(3.56\pm0.18)\times 10^{-9}$,
and discuss the error budget.
\end{abstract}

\vspace*{0.5truecm}
\vfill
\noindent
{\rm March 2013}
\vspace*{0.5truecm}

\end{titlepage}

\thispagestyle{empty}
\vbox{}
\newpage

\setcounter{page}{1}

\section{Introduction}\label{sec:intro}
The rare decay $\Bsmumu$ is very strongly suppressed within the Standard Model (SM).
As this need not be the case for SM extensions, which could dramatically enhance it, it has already for decades played an important role in constraining such extensions and giving hope for seeing a clear signal of New Physics (NP) \cite{Buras:2012ts}. 
Using the parametric formula for its branching ratio from Ref.~\cite{Buras:2012ru}, and updating the values for the $B_s$-decay constant $F_{B_s}$ \cite{Dowdall:2013tga} and the $B_s$-lifetime $\tau_{B_s}$~\cite{Amhis:2012bh},
we find~\footnote{This should be compared with ${\rm BR}(\Bsmumu)_{\rm SM} = (3.23\pm 0.27)\times 10^{-9}$ in Ref.~\cite{Buras:2012ru}, implying that the central value remains practically  unchanged but the error has decreased significantly. We discuss the reason for this change in Subsection~\ref{BRDiscussion}.}
\begin{equation}
    {\rm BR}(\Bsmumu)_{\rm SM} = (3.25\pm 0.17)\times 10^{-9}.
\label{BRtheoRes}
\end{equation}
Thus, within the SM, only about one in every 300 million $B_s^0$ mesons is predicted to decay to a pair of muons. 

Concerning the measurement of this branching ratio, a complication arises due to 
the presence of $B^0_s$--$\bar B^0_s$ oscillations \cite{deBruyn:2012wk}. 
In particular, LHCb has recently established a sizable value of the decay 
width difference $\Delta\Gamma_s$ between the
$B_s$ mass eigenstates  \cite{Raven:2012fb}:
\begin{equation}\label{defys}
	y_s\equiv\frac{\Gamma_{\rm L}^{(s)} - \Gamma_{\rm H}^{(s)}}{\Gamma_{\rm L}^{(s)} + \Gamma_{\rm H}^{(s)}}=
\frac{\Delta\Gamma_s}{2\,\Gamma_s}
=0.087\pm0.014,
\end{equation}
where $\Gamma_s = \tau_{B_s}^{-1}$ denotes the average $B_s$ decay width. This quantity enters the
time-integrated decay rate, which is at the origin of the measurement of the ``experimental"
branching ratio $\overline{\rm BR}(B_{s}\to\mu^+\mu^-)$. 
It is related to the ``theoretical" branching ratio, referring to $y_s=0$, as follows~\cite{deBruyn:2012wk}:
\begin{equation}\label{BR-conv}
{\rm BR}(B_{s}\to\mu^+\mu^-)=\left[
\frac{1-y_s^2}{1+\mathcal{A}^{\mu\mu}_{\Delta\Gamma} y_s} \right]
\overline{\rm BR}(B_{s}\to\mu^+\mu^-),
\end{equation}
where the observable $\mathcal{A}^{\mu\mu}_{\Delta\Gamma}$ equals $+1$ in the SM. 
Using the numerical value for the SM branching ratio in (\ref{BRtheoRes}) and the
experimental value for $y_s$ in (\ref{defys}) gives
\be
\label{FleischerSM}
\overline{\rm BR}(B_{s}\to\mu^+\mu^-)_{\rm SM}= (3.56\pm0.18)\times 10^{-9},
\ee
which is the reference value for comparing the time-integrated experimental  
branching ratio with the SM.

Over the last decade we have seen the upper bounds for the $\Bsmumu$ branching ratio
continuously move down thanks to the CDF and D0 collaborations at the Tevatron and
the ATLAS, CMS and LHCb experiments at the LHC (for a review, see Ref.~\cite{Albrecht:2012hp}). 
In November 2012, the LHCb collaboration reported the first evidence for
the $\Bsmumu$ decay at the $3.5\sigma$ level, with the following branching ratio 
\cite{LHCbBsmumu}:
\begin{equation}\label{BRexpMeas}
	\overline{\rm BR}(\Bsmumu)_\text{LHCb} = (3.2^{+1.5}_{-1.2})\times 10^{-9}
	\, \in [1.1,6.4]\times 10^{-9} \, (95\%\ {\rm C.L}.).
	\end{equation}
The agreement with (\ref{FleischerSM}) is remarkable, although the rather large 
experimental error still allows for sizable NP contributions. It is already obvious at this 
stage, however, that this will be a challenging endeavour.

As emphasized in Ref.~\cite{deBruyn:2012wk}, and discussed and illustrated in 
detail below, the observable $\mathcal{A}^{\mu\mu}_{\Delta\Gamma}$ entering 
(\ref{BR-conv}) is sensitive to NP contributions. 
It is thus a complementary observable to the $\Bsmumu$ branching ratio,
offering independent information on the short-distance structure of this decay.
The observable $\mathcal{A}^{\mu\mu}_{\Delta\Gamma}$ can be extracted from the untagged data sample, for which no distinction is made between initially present $B^0_s$ or 
$\bar B^0_s$ mesons, once enough decay events are available for the decay-time information to be accurately taken into account\cite{deBruyn:2012wj,deBruyn:2012wk}. 
The conversion factor in (\ref{BR-conv}) would then be determined from the data, thereby also allowing the extraction of the theoretical $\Bsmumu$ branching ratio.

If tagging information is included, requiring even more events to compensate the efficiency of distinguishing between initially present $B^0_s$ or 
$\bar B^0_s$ mesons, a CP-violating, time-dependent rate 
asymmetry can be measured. As we assume that the muon helicity will not be measured, 
this rate asymmetry is governed by another, third observable,  ${\cal S}_{\mu\mu}$, which 
vanishes in the SM but is very sensitive to new CP-violating effects entering 
$\Bsmumu$ in extensions of the SM \cite{deBruyn:2012wk}. Analyses of CP-violating effects
in $B_{s(d)}\to\ell^+\ell^-$ decays, which neglected $\Delta\Gamma_s$ effects, were performed for
various models of NP in 
Refs.~\cite{Huang:2000tz,Huang:2002dj,Dedes:2002er,Chankowski:2004tb}. 
An analysis of $Z'$ models that including $\Delta\Gamma_s$ effects was performed in Ref.~\cite{Buras:2012jb}.

The observables $\mathcal{A}^{\mu\mu}_{\Delta\Gamma}$ and ${\cal S}_{\mu\mu}$ both depend
on the CP-violating $B^0_s$--$\bar B^0_s$ mixing phase 
\begin{equation}
\phi_s=\phi_s^{\rm SM}+\phi_s^{\rm NP} = -2\lambda^2\eta + \phi_s^{\rm NP},
\end{equation}
where the numerical value of the SM piece, involving the Wolfenstein parameters 
$\lambda$ and $\eta$ of the CKM matrix, is given by $-(2.08\pm0.09)^\circ$. 
The LHCb analysis of CP-violation in the $B_s\to J/\psi \phi$ decay currently gives the most precise experimental determination of this phase~\cite{Raven:2012fb}:
\footnote{We avoid averages for $\phi_s$ that include decays such as $B^0_s\to J/\psi f_0(980)$ because of the unsettled hadronic structure of the $f_0(980)$ state \cite{Fleischer:2011au}.}
\begin{equation}\label{phis}
\phi_s=-\left(0.06 \pm 5.99\right)^\circ.
\end{equation}
Hadronic uncertainties from doubly Cabibbo-suppressed penguin effects have been neglected in this measurement; these corrections have to be controlled once the experimental precision improves further \cite{Fleischer:2012dy}.

As the measurements of the observables $\mathcal{A}^{\mu\mu}_{\Delta\Gamma}$ and 
${\cal S}_{\mu\mu}$ refer to the era of the LHCb upgrade, we can assume that $\phi_s$
will be known precisely once data for these observables become available. The goal of the present 
paper is to investigate and illustrate how the experimental knowledge of the trio from 
$\Bsmumu$,
\be\label{trio}
\overline{\rm BR}(\Bsmumu), \qquad \mathcal{A}^{\mu\mu}_{\Delta\Gamma}, 
\qquad {\cal S}_{\mu\mu},
\ee
will shed light on the possible presence of NP in this decay, which 
cannot be obtained on the basis of the information on $\overline{\rm BR}(\Bsmumu)$ alone.

Our paper is organized as follows: in Section~\ref{sec:obs} we recall the 
definitions of the observables in (\ref{trio}) and discuss their 
various properties. In Section~\ref{sec:scenarios} we introduce various 
scenarios for NP, classifying them in terms of four general parameters 
which can in principle be calculated in any fundamental model and 
are directly related to the physics of $\Bsmumu$.
In this context we classify a selection of popular 
NP models into the considered scenarios.
In Section~\ref{Trees}
we consider three classes of specific  NP models. The first one with  tree level neutral gauge boson contributions to FCNC processes that could represent $Z'$ models or models with flavour violating $Z$ couplings. The second 
one in which the role of gauge bosons is taken over by scalar or pseudoscalar 
tree-level exchanges. Finally, we present
a third model in which both a scalar and a pseudoscalar with the same mass couple equally to quarks and leptons.
We demonstrate how the measurements of the 
observables in (\ref{trio}) can 
distinguish between these three classes of models. In Section~\ref{Summary} we 
summarize the highlights of our paper.

\section{Observables of $\boldsymbol \Bsmumu$}\label{sec:obs}
\subsection{Basic Effective Hamiltonian}

The general model-independent low-energy effective Hamiltonian for a $B_s \to \ell^+ \ell^-$ decay is \cite{Altmannshofer:2011gn,Beaujean:2012uj}
\begin{equation}
	{\cal H}_{\rm eff} = -\frac{G_F\,\alpha}{\sqrt{2}\pi}\left\{
		V_{ts}^* V_{tb}\,  \sum_{i}^{10,S,P} \left( C_i\,{\cal O}_i + C'_i\,{\cal O}'_i\right) 
		+ {\rm h.c}\right\},
		\label{effOPE}
\end{equation}
where the operators are 
\begin{align}
	{\cal O}_{10} &= (\bar s\gamma_\mu P_L b)(\bar l \gamma^\mu \gamma_5 l), & 
	{\cal O}'_{10} &= (\bar s\gamma_\mu P_R b)(\bar l \gamma^\mu \gamma_5 l), \notag\\	
	{\cal O}_{S} &= m_b (\bar s P_R b)(\bar l l), &
	{\cal O}'_{S} &= m_b (\bar s P_L b)(\bar l l), \notag\\	
	{\cal O}_{P} &= m_b (\bar s P_R b)(\bar l \gamma_5 l), & 
	{\cal O}'_{P} &= m_b (\bar s P_L b)(\bar l \gamma_5 l)
\end{align}
and $\alpha=e^2/4\pi$ is the QED fine structure constant.
The observables that we will calculate below can each be expressed in terms of the following combinations of Wilson Coefficients: 
\begin{align}
	P &\equiv \frac{C_{10} - C'_{10}}{C_{10}^{\rm SM}} + \frac{m_{B_s}^2}{2m_\mu}\left(\frac{m_b}{m_b + m_s}\right) \left( \frac{C_{P} - C'_{P}}{C_{10}^{\rm SM}}\right)\equiv |P|e^{i\varphi_P}, \notag\\
	S &\equiv \sqrt{1 - \frac{4\,m_\mu^2}{m_{B_s}^2}}\frac{m_{B_s}^2}{2m_\mu}\left(\frac{m_b}{m_b + m_s}\right) \left( \frac{C_{S} - C'_{S}}{C_{10}^{\rm SM}}\right)\equiv |S|e^{i\varphi_S}.
	\label{PSdefn}
\end{align}
In the SM $C_{10}=C_{10}^{\rm SM}$ and $C'_{10}$, $C_S^{(\prime)}$ and $C_P^{(\prime)}$ are all negligibly small, so that $P^{\rm SM}=1$ and $S^{\rm SM}=0$.
The Wilson coefficient $C_{10}$ is given in the SM as follows 
\begin{equation}
	C_{10}^{\rm SM} =-\eta_Y  \sin^{-2}\theta_WY_0(x_t)=-4.134,
\end{equation}
where $Y_0(x_t)$ is a one-loop function with $x_t=m_t^2/M_W^2$ \cite{Buchalla:1990qz},
and the coefficient $\eta_{Y}$ is a QCD factor that for $m_t=m_t(m_t)$ is close to unity: $\eta_{Y}=1.012$
\cite{Buchalla:1998ba,Misiak:1999yg}.
The scalar Wilson coefficients are thereby related to the parameter $S$ by the numerical factor 
\begin{equation}
	m_b(C_{S} - C'_{S}) = -0.130 \times S.
\end{equation}
Note that while $C_{10}$ and $C_{10}'$ are dimensionless, the coefficients 
 $C_S^{(\prime)}$ and $C_P^{(\prime)}$ have dimension $\gev^{-1}$.

\subsection{Time-Dependent Rates}

The time-dependent rate for a $B^0_s$ meson decaying to two muons with a specific helicity $\lambda= L,R$ is given by
\begin{align}
	\Gamma(B_s^0(t)\to \mu_\lambda^+ \mu_\lambda^-)
	=& \frac{G_F^4\, M_W^4\sin^4\theta_W}{16\pi^5}\left|C_{10}^{\rm SM} V_{ts} V_{tb}^*\right|^2 F_{B_s}^2 m_{B_s} m_\mu^2\ 
	\sqrt{1- \frac{4m^2_\mu}{m^2_{B_s}}}\times\left(|P|^2 + |S|^2\right)\,\notag\\
	&\times \bigg\{ {\cal C}_{\mu\mu}^\lambda\cos(\Delta M_s\,t) + {\cal S}_{\mu\mu}\cos(\Delta M_s\,t) \notag\\
	&\quad\quad + \cosh\left(\frac{y_s\,t}{\tau_{B_s}}\right) + {\cal A}_{\Delta\Gamma}^{\mu\mu}\sinh\left(\frac{y_s\,t}{\tau_{B_s}}\right)\bigg\}\times\,e^{-t/\tau_{B_s}},
	\label{timeDepHelRate}
\end{align}
where $\tau_{B_s}\equiv 2/(\Gamma_{\rm H} + \Gamma_{\rm L})$ is the $B_s$ mean lifetime and $y_s$ is defined in (\ref{defys}).

The time-dependent rate for a $\bar B^0_s$ meson is obtained from the above expression by replacing ${\cal C}_{\mu\mu}^\lambda\to - {\cal C}_{\mu\mu}^\lambda$ and ${\cal S}_{\mu\mu}\to -{\cal S}_{\mu\mu}$.
The time-dependent observables for both rates can be expressed in terms of the parameters defined in \eqref{PSdefn} as~\cite{deBruyn:2012wk,Fleischer:2012fy}
\begin{align}
	{\cal C}_{\mu\mu}^\lambda &=  -\eta_\lambda\left[\frac{2|PS|\cos(\varphi_P-\varphi_S)}{|P|^2+|S|^2} \right],\label{Cobs}\\
	{\cal S}_{\mu\mu}
	&=\frac{|P|^2\sin(2\varphi_P-\phi_s^{\rm NP})-|S|^2\sin(2\varphi_S-\phi_s^{\rm NP})}{|P|^2+|S|^2},\\
	{\cal A}^{\mu\mu}_{\Delta\Gamma} &= \frac{|P|^2\cos(2\varphi_P-\phi_s^{\rm NP}) - |S|^2\cos(2\varphi_S-\phi_s^{\rm NP})}{|P|^2 + |S|^2}.
	\label{ADG}
\end{align}
The phase $\phi_s^{\rm NP}$ represents the CP-violating 
NP contributions to $B_s^0$--$\bar B_s^0$ mixing. 
It influences the mixing-induced CP asymmetry in the $B_s^0(\bar B_s^0)\to\psi\phi$ decays
\cite{Dunietz:2000cr}, with the latter given by 
\begin{equation}
S_{\psi\phi} = -\sin\phi_s= \sin(2|\beta_s|-\phi_s^{\rm NP})\,,\qquad V_{ts}=-|V_{ts}| e^{-i\beta_s}
\label{eq:3.44}
\end{equation}
with $\beta_s\simeq -1^\circ\,$. In Ref.~\cite{Buras:2012ts} and the papers 
reviewed there $\phi_s^{\rm NP}=2\varphi_{B_s}$.

Only the observable ${\cal C}_{\mu\mu}^\lambda$ is dependent on the helicity of the final state i.e.\ it depends on the parameter $\eta_\lambda \equiv \{\mbox{+1:}~L;-\mbox{1:}~R\}$.
The presence of the observable $\mathcal{A}^{\mu^+\mu^-}_{\Delta\Gamma}$ is a consequence of the sizable $B_s$ decay width difference $\Delta\Gamma_s$.

In practice the muon helicities $\lambda$ are very challenging to measure.
If no attempt is made to disentangle them, then we measure their sum:
\begin{align}
	\Gamma(B_s^0(t)\to \mu^+ \mu^-) &\equiv \sum_{\lambda=L,R} \Gamma(B_s^0(t)\to \mu_\lambda^+ \mu_\lambda^-),\notag\\
	\Gamma(\bar B_s^0(t)\to \mu^+ \mu^-) &\equiv \sum_{\lambda=L,R} \Gamma(\bar B_s^0(t)\to \mu_\lambda^+ \mu_\lambda^-).
\end{align}
Observe from equations \eqref{timeDepHelRate} and \eqref{Cobs} that ${\cal C}_{\mu\mu}^\lambda$, which was dependent on the muon helicity, cancels in both sums~\cite{deBruyn:2012wk}.

The $\Bsmumu$ helicity-summed time-dependent untagged rate is then given by
\begin{align}
	\langle\Gamma(B_s(t)\to \mu^+ \mu^-)\rangle 
	\equiv&\ \Gamma(B_s^0(t)\to \mu^+ \mu^-) + \Gamma(\bar B_s^0(t)\to \mu^+ \mu^-) \notag\\
	=&\ \frac{G_F^4\, M_W^4\sin^4\theta_W}{4\pi^5}\left|C_{10}^{\rm SM} V_{ts} V_{tb}^*\right|^2 F_{B_s}^2 m_{B_s} m_\mu^2\ 
	\sqrt{1- \frac{4m^2_\mu}{m^2_{B_s}}}\notag\\
	&\ \times\,\left(|P|^2 + |S|^2\right) \notag\\
	&\ \times\,e^{-t/\tau_{B_s}}\left[\cosh\left(y_s\, t/\tau_{B_s}\right) + {\cal A}^{\mu\mu}_{\Delta\Gamma}\sinh\left(y_s\, t/\tau_{B_s}\right)\right].
	\label{untaggedRate}
\end{align}
Similarly, the helicity-summed time-dependent tagged rate asymmetry is 
\begin{align}
\frac{\Gamma(B^0_s(t)\to \mu^+\mu^-)-
\Gamma(\bar B^0_s(t)\to \mu^+\mu^-)}{\Gamma(B^0_s(t)\to \mu^+\mu^-)+
\Gamma(\bar B^0_s(t)\to \mu^+\mu^-)}
=\frac{{\cal S}_{\mu\mu}\sin(\Delta M_st)}{\cosh(y_st/ \tau_{B_s}) + 
{\cal A}^{\mu\mu}_{\Delta\Gamma} \sinh(y_st/ \tau_{B_s})}.
\label{summedRate}
\end{align}
It is important to clarify that although there is no explicit term for direct CP violation in the rate asymmetry, 
this does not mean that the absolute values squared of ${\cal S}_{\mu\mu}$ and ${\cal A}^{\mu\mu}_{\Delta\Gamma}$ necessarily sum to one.
These two observables also have an implicit dependence on ${\cal C}^\lambda_{\mu\mu}$, the rate asymmetry for $B_s^0$ and $\bar B_s^0$ decays to the specific helicity muon final states.
This gives the relation
\begin{equation}
	| {\cal S}_{\mu\mu}|^2 + |{\cal A}^{\mu\mu}_{\Delta\Gamma}|^2 = 1 - |{\cal C}^\lambda_{\mu\mu}|^2 = 1 - \left[\frac{2|PS|\cos(\varphi_P - \varphi_S)}{|P|^2 + |S|^2}\right]^2.
	\label{obsDependence}
\end{equation}
Thus if there are no new CP-violating phases in the mixing or decay amplitudes, $\varphi_P=\varphi_S=\phi_s^{\rm NP}=0$ such that ${\cal S}_{\mu\mu}=0$, ${\cal A}^{\mu\mu}_{\Delta\Gamma}$ does not have to take its SM value of 1.
The presence of a non-negligible scalar operator ${\cal O}^{(\prime)}_S$, so that $|S|\neq 0$, is sufficient to ensure that ${\cal A}^{\mu\mu}_{\Delta\Gamma}\neq 1$, as can also be seen from (\ref{ADG}).

In contrast to the branching ratio, the dependence on $F_{B_s}$ cancels in both 
$\mathcal{A}^{\mu\mu}_{\Delta\Gamma}$ and ${\cal S}_{\mu\mu}$, and these observables
are also not affected by CKM uncertainties. Consequently, they are theoretically 
clean.\footnote{In principle, corrections arise from loop topologies with internal charm- and 
up-quark exchanges. However, these are strongly suppressed by the CKM ratio 
$|V_{us}^*V_{ub}/V_{ts}^*V_{tb}|\sim 0.02$, and are even further suppressed dynamically
for $B_s\to\mu^+\mu^-$. These effects do hence not play any role for these observables 
from the practical point of view.}  
Moreover, these observables are also not affected by the ratio $f_d/f_s$ of fragmentation functions, which are the major limitation of the precision of the $B_s\to\mu^+\mu^-$ branching ratio measurement at hadron colliders~\cite{Fleischer:2010ay}.
As $\mathcal{A}^{\mu\mu}_{\Delta\Gamma}$ does not rely on flavour tagging, which is difficult for a rare decay, it will be easier to determine.
Given enough statistics, a full fit to the time-dependent untagged rate will give ${\cal A}^{\mu\mu}_{\Delta\Gamma}$.
With limited statistics, an {\it effective lifetime} measurement may be easier, which corresponds to fitting a single exponential to this rate. For a maximal likelihood fit, the $\Bsmumu$ effective lifetime is equal to the time expectation value of \eqref{untaggedRate} \cite{deBruyn:2012wj}: 
\begin{equation}
	\tau_{\mu\mu} \equiv \frac{\int^\infty_0 t\,\langle\Gamma(B_s(t)\to \mu^+ \mu^-)\rangle\, dt}{\int_0^\infty \langle\Gamma(B_s(t)\to \mu^+ \mu^-)\rangle\, dt}.
\end{equation}
The untagged observable is then given by
\begin{equation}
	{\cal A}^{\mu\mu}_{\Delta\Gamma} = \frac{1}{y_s}\left[\frac{(1-y_s^2)\,\tau_{\mu\mu} - (1+y_s^2)\tau_{B_s}}{2\tau_{B_s} - (1-y_s^2)\,\tau_{\mu \mu}}\right].
\end{equation}

\subsection{The Branching Ratio}\label{BRDiscussion}

A $\Bsmumu$ branching ratio measurement amounts to counting all events over all (accessible) time, 
and is thus defined as the time integral of the untagged rate given in \eqref{untaggedRate}~\cite{Dunietz:2000cr,deBruyn:2012wj,deBruyn:2012wk}:
\begin{align}
	\overline{\rm BR}(\Bsmumu) \equiv \frac{1}{2}\int_0^\infty \langle\Gamma(B_s(t)\to \mu^+ \mu^-)\rangle\,dt.
	\label{BRexp}
\end{align}

LHCb has recently presented the first measurement of the $\Bsmumu$ {\it time-integrated} rate~\cite{LHCbBsmumu} that we have given in (\ref{BRexpMeas}).
In contrast, the SM prediction for the $\Bsmumu$ branching ratio in 
(\ref{BRtheoRes})
is computed theoretically for one instant in time, namely at $t=0$ 
i.e.\ it neglects the effects of $B_s^0$--$\bar B_s^0$ mixing.
Specifically, it is given by
\begin{align}
	{\rm BR}(\Bsmumu)_{\rm SM} &= \frac{\tau_{B_s}}{2}\langle\Gamma(B_s(t)\to \mu^+ \mu^-)\rangle\Big|_{t=0,\,P=1,\,S=0}\notag\\
	&=  \frac{\tau_{B_s}\,G_F^4\, M_W^4\sin^4\theta_W}{8\pi^5}\left|C_{10}^{\rm SM} V_{ts} V_{tb}^*\right|^2 F_{B_s}^2 m_{B_s} m_\mu^2\ \sqrt{1- \frac{4m^2_\mu}{m^2_{B_s}}};
	\label{BRSM}
\end{align}
an updated numerical estimate  is given in (\ref{BRtheoRes}).

It is now straightforward to derive the expression
\be
\frac{{\rm BR}(\Bsmumu)}{{\rm BR}(\Bsmumu)_{\rm SM}}
	= |P|^2 + |S|^2.
\ee
However, as not the theoretical but the experimental branching ratio is measured, it is useful
to introduce the following ratio \cite{deBruyn:2012wk}:
\begin{align}
    \overline{R} &\equiv \frac{\overline{\rm BR}(\Bsmumu)}{\overline{\rm BR}(\Bsmumu)_{\rm SM}}
	= \left[\frac{1+{\cal A}^{\mu\mu}_{\Delta\Gamma}\,y_s}{1+y_s} \right] \times (|P|^2 + |S|^2)\notag\\
	&= \left[\frac{1+y_s\cos(2\varphi_P-\phi_s^{\rm NP})}{1+y_s} \right] |P|^2 + \left[\frac{1-y_s\cos(2\varphi_S-\phi_s^{\rm NP})}{1+y_s} \right] |S|^2,
\end{align}
where the sizable decay width difference $\Delta\Gamma_s$ enters.
The parameter $\overline{R}$ is related to $R$ defined in Ref.~\cite{deBruyn:2012wk} by $\overline{R} = (1-y_s) R$.
Combining the theoretical SM prediction in \eqref{BRtheoRes} with the experimental
result in \eqref{BRexpMeas} gives
\begin{equation}
    \overline{R} = 0.90^{+0.42}_{-0.34} 
	\, \in [0.30,1.80] \,  (95\%\ {\rm C.L}).
    \label{RexpValues}
\end{equation}
This range should be compared with the SM value $\overline{R}_{\rm SM}=1$.

Finally we would like to explain the origin of the reduced error in 
(\ref{BRtheoRes}). To this end we return to the basic parametric formula (18) for the theoretical branching ratio in Ref.~\cite{Buras:2012ru}. It turns out that the changes of the input parameters over the last six months have practically no impact on the central value obtained there. Indeed updating the central values of $F_{B_s}$ and $\tau_{B_s}$, we 
can cast this formula into the following expression:
\begin{equation}
{\rm BR}(\Bsmumu)_{\rm SM} = 3.25\times 10^{-9} \left(\frac{M_t}{173.2 \gev}\right)^{3.07}\left(\frac{F_{B_s}}{225\mev}\right)^2
\left(\frac{\tau_{B_s}}{1.500 {\rm ps}}\right)\left|\frac{V_{tb}^*V_{ts}}{0.0405}\right|^2.
\label{BRtheoRpar}
\end{equation}
The most recent world averages  for $F_{B_s}$ \cite{Dowdall:2013tga} and $\tau_{B_s}$  \cite{Amhis:2012bh} are 
\be
F_{B_s}=(225\pm 3)~\mev, \qquad  \tau_{B_s}=1.503 (10) ~{\rm ps}
\ee
to be compared with $F_{B_s}=(227\pm 8)~\mev$ and  $\tau_{B_s}=1.466 (30) ~{\rm ps}$ used in Ref.~\cite{Buras:2012ru}. While the change in 
$\tau_{B_s}$ is an experimental improvement,  confirmation of the impressive accuracy on  $F_{B_s}$ is eagerly awaited. In 
Ref.~\cite{Buras:2012ru} a more 
conservative approach has been used, but here  we follow Ref.~\cite{Dowdall:2013tga},
updating also $\tau_{B_s}$. With unchanged input on $M_t$ and $V_{ts}$ with 
respect to Ref.~\cite{Buras:2012ru} we arrive at (\ref{BRtheoRes}) and 
consequently, after including the correction from $\Delta\Gamma_s$, at 
(\ref{FleischerSM}).

\begin{figure}[t]
\begin{center}
\includegraphics[height=5.5cm]{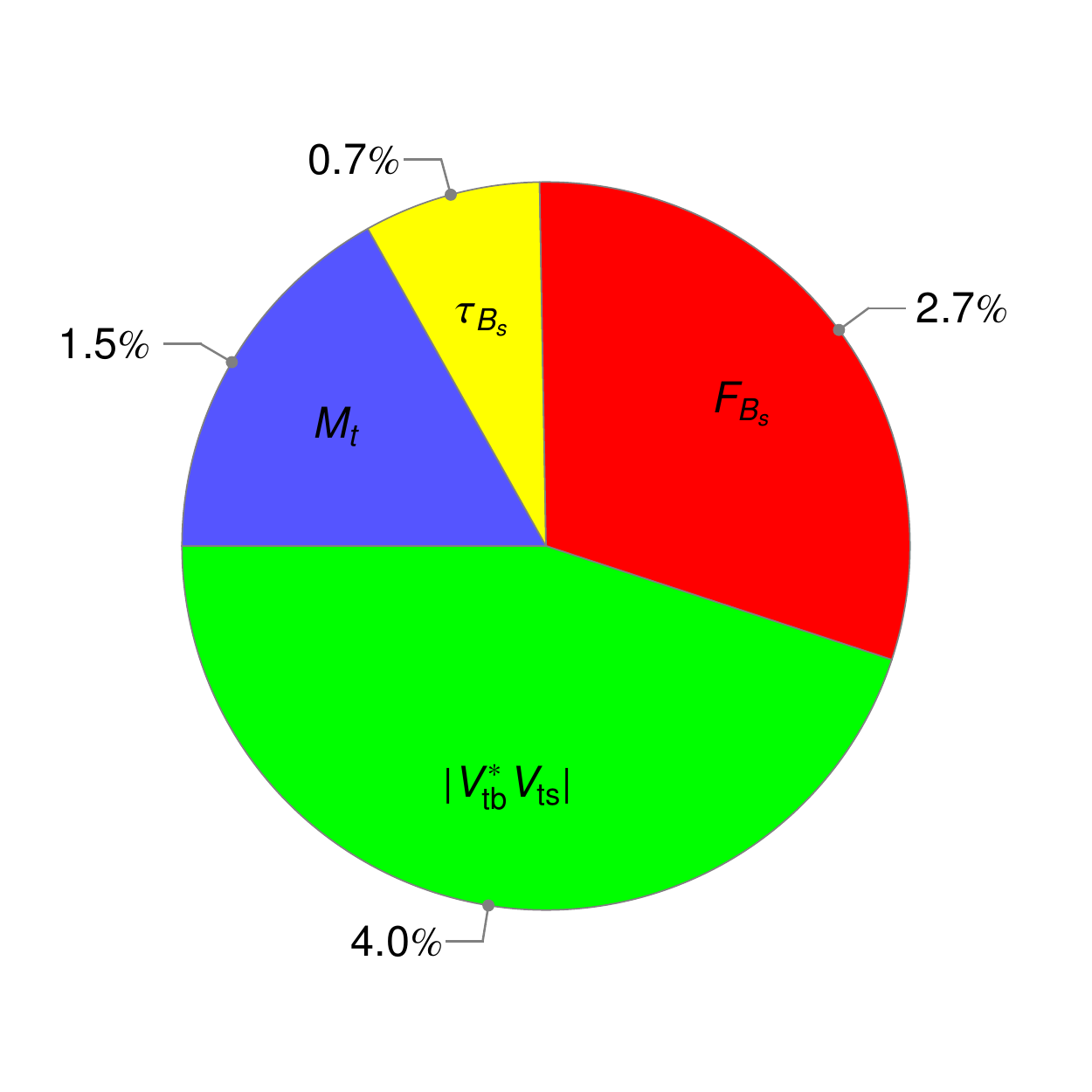}
\includegraphics[height=5.5cm]{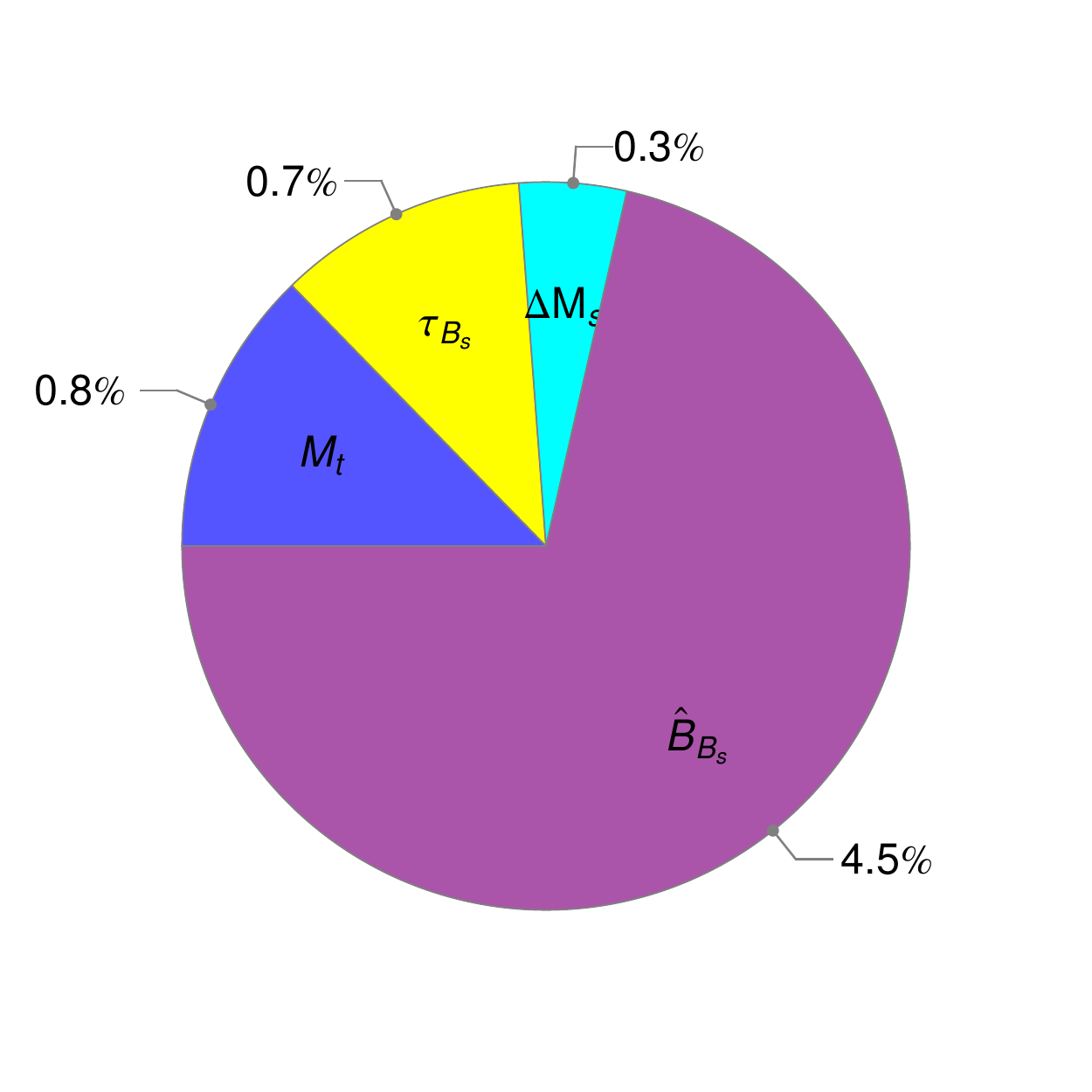}
\end{center}
\caption{Error budgets for the two branching ratio calculations of $\Bsmumu$ in the Standard Model given in \eqref{BRtheoRpar} (left) and \eqref{BRAJB} (right).}
\label{fig:pies}
\end{figure}

Now as stressed and analysed in  \cite{Buras:2012ru,Misiak:2011bf} additional modifications 
could come from complete NLO electroweak corrections, which have just been completed 
(M. Gorbahn, private communication) and affect the overall factor in (\ref{BRtheoRpar}) 
by roughly $3\%$. The leftover uncertainties due to unknown NNLO corrections are therefore 
fully negligible. Taking at face value the present error 
on $F_{B_s}$, the current error budget for the branching ratio is as follows:
\be\label{budget}
M_t:~1.5\%, \qquad  F_{B_s}: 2.7\%, \qquad \tau_{B_s}:~ 0.7\%, \qquad 
|V_{tb}^*V_{ts}|:~4\%,
\ee
 It is also depicted in the left panel of Figure~\ref{fig:pies}. Evidently, after completion of 
 NLO electroweak effects and improved values of $F_{B_s}$, the error on
$|V_{tb}^*V_{ts}|$ is now the largest uncertainty  but this assumes that the error on  
$F_{B_s}$ is indeed as small as obtained 
in Ref.~\cite{Dowdall:2013tga}.

While the small error on  $F_{B_s}$ is expected to be consolidated soon, the 
decrease of the error in $|V_{ts}|$ appears to be much harder. In this context
it should be recalled that the branching ratio in question can also be 
calculated by using the mass difference $\Delta M_s$ \cite{Buras:2003td}. The 
updated parametric formula (13) of the latter paper reads 
\begin{equation}
{\rm BR}(\Bsmumu)_{\rm SM} = 3.38\times 10^{-9} \left(\frac{M_t}{173.2 \gev}\right)^{1.6} \left(\frac{\tau_{B_s}}{1.500 {\rm ps}}\right)
\left(\frac{1.33}{\hat B_{B_s}}\right)
\left(\frac{\Delta M_s}{17.72/{\rm ps}}\right).
\label{BRAJB}
\end{equation}
We note that 
among the uncertainties in (\ref{budget}), the largest 
two are absent and the uncertainty due to $M_t$ is reduced to $0.8\%$. 
The uncertainty due to $\Delta M_s$ is negligible \cite{Amhis:2012bh}. The 
error budget for this expression reads 
\be\label{budget1}
M_t:~0.8\%, \qquad  \tau_{B_s}:~ 0.7\%, \qquad \Delta M_s: ~0.3\%, \qquad \hat B_{B_s}:~4.5\%,
\ee
where we used $\hat B_{B_s}=1.33(6)$ \cite{Laiho:2009eu}. 
 It is pictured in the right panel of Figure~\ref{fig:pies}.
The latter uncertainty is expected to be reduced significantly in the coming years so that 
(\ref{BRAJB}) could remain to be the most accurate estimate of the branching 
ratio in question under the assumption that there are no NP contributions to 
$\Delta M_s$ and the SM reproduces its experimental value. In view of this 
tacit assumption in (\ref{BRAJB}), we prefer to use (\ref{BRtheoRes}) as the 
present best estimate of the theoretical branching ratio. On the other hand, by using (\ref{BRAJB}) we find, after the inclusion of $\Delta\Gamma_s$ the 
correction,
\begin{equation}
\overline{\rm BR}(\Bsmumu)_{\rm SM} = (3.71\pm 0.17)\times 10^{-9}, \qquad (\Delta M_s)
\label{BRAJBfinal}
\end{equation}
which agrees very well with  (\ref{FleischerSM}). This updates the 
estimate  $(3.2\pm 0.2)\times 10^{-9}$ in Ref.~\cite{Buras:2010wr},
where  $\Delta\Gamma_s$ effects where not included and other input, 
in particular the value of $\tau_{B_s}$, was different than now.

\section{Constrained Scenarios and Their Phenomenology}\label{sec:scenarios}
\subsection{Preliminaries}
Experiments have started honing in on the $\Bsmumu$ time-integrated rate, or branching ratio, for which the observable $\overline{R}$ parameterises possible NP contributions.
Next in line is a time-dependent analysis, first without tagging, giving ${\cal A}^{\mu\mu}_{\Delta\Gamma}$, and then with tagging, giving ${\cal S}_{\mu\mu}$.
The end result will be three experimental observables, which, if there are scalar operators contributing to the decay mode, can each contain independent information (see the discussion around \eqref{obsDependence}).

With the phase $\phi_s^{\rm NP}$ already significantly constrained 
by the current data (\ref{phis}), these three observables depend on four unknowns: 
\be\label{unknowns}
|P|,\qquad \varphi_P, \qquad |S|, \qquad \varphi_S. 
\ee
Therefore  we cannot in general solve for all of these model-independent NP parameters by considering the decay $B_s\to\mu^+\mu^-$ alone.
One solution is to invoke other $b\to s \mu^+\mu^-$  transitions like 
the decays $B\to K \mu^+\mu^-$  and $B\to K^*\mu^+\mu^-$. In particular, as emphasized in Ref.~\cite{Becirevic:2012fy}, observables in $B\to K \mu^+\mu^-$ are sensitive to $C_{S,P}+C^\prime_{S,P}$, rather then differences of these coefficients, thereby allowing additional complementary tests and in principle the determination 
of all Wilson coefficients. But present form factor uncertainties in these 
decays do not yet provide significant new constraints on scalar operators 
relatively to the ones obtained from $B_s\to\mu^+\mu^-$. In any case 
such analysis would be beyond 
the scope of the present paper.

In the spirit of the analysis  of $Z^\prime$ contributions to FCNC 
processes in Ref.~\cite{Buras:2012jb}, where various scenarios for  $Z^\prime$ 
couplings to quarks have been considered, and an analogous analysis for tree-level 
scalar and pseudoscalar contributions to FCNC processes \cite{Buras:2013rqa}, 
we will consider various scenarios for $S$ and $P$ that 
will allow us to reduce the number of free NP parameters and eventually, 
with the help of future data, uniquely determine them. 
Our scenarios are motivated by generic features of NP models and, as we will 
see below, they result in a distinct phenomenology for the observables $\overline{R}$, ${\cal A}^{\mu\mu}_{\Delta\Gamma}$ and ${\cal S}_{\mu\mu}$. 
In the present section our analysis is dominantly phenomenological, 
although we discuss the motivation behind each scenario and the characteristic features of its phenomenology.
Moreover we indicate what kind of fundamental physics could be at the basis of each scenario considered and we survey specific models of NP and categorise them into the scenarios that we will list now.

The five scenarios to be considered are as follows:
\begin{enumerate}[(A)]
	\item $S=0$
	\item $P= 1$
	\item $P\pm S= 1$
	\item $\varphi_P,\varphi_S\in \{0,\pi\}$
	\item $P = 0$.
\end{enumerate}
The scenarios are intended to be limiting cases, i.e.\ although we are not aware of a model that predicts $P=0$, $P\approx 0$ is conceivable and the resulting phenomenology will be approximately the same.

\subsection{Scenario A: $\boldsymbol{S=0}$}
\subsubsection{General Formulation}

This scenario is realised if $C_S-C'_S=0$,
leaving $C^{(\prime)}_{10}$ and $C^{(\prime)}_P$ free to take non-SM values as well as 
CP-violating phases.
Thus models with only new gauge bosons or pseudoscalars naturally fall into this category and consequently, as we will see below, this scenario 
includes a number of popular BSM models.
Also models with scalars can qualify, provided the scalars couple left-right symmetrically to quarks so that $C_S=C'_S$.

In this scenario the rate asymmetry between $B_s^0$ and $\bar B_s^0$ decays to the individual muon helicities vanishes: ${\cal C}_{\mu\mu}^\lambda=0$.
Therefore the two time-dependent observables do not carry independent information, being bound by the constraint
\begin{equation}
	|{\cal S}_{\mu\mu}|^2 + |{\cal A}^{\mu\mu}_{\Delta\Gamma}|^2 = 1.
\end{equation}
Specifically, 
\begin{align}\label{FA1}
	{\cal A}^{\mu\mu}_{\Delta\Gamma} = \cos(2\varphi_P - \phi_s^{\rm NP}),\quad
	{\cal S}_{\mu\mu} = \sin(2\varphi_P - \phi_s^{\rm NP}),
\end{align}
while the branching ratio observable is given by
\begin{equation}\label{FA2}
	\overline{R} = |P|^2 \left[\frac{1+y_s\cos(2\varphi_P - \phi_s^{\rm NP})}{1+y_s}\right].
\end{equation}

Formulae (\ref{FA1}) and (\ref{FA2}) are the basic expressions for this scenario. 
The three observables in  (\ref{FA1}) and (\ref{FA2}) are given in terms of two 
unknowns: $|P|$ and $\varphi_P$. As we assume knowledge of $\phi_s^{\rm NP}$,
${\cal A}^{\mu\mu}_{\Delta\Gamma}$ and ${\cal S}_{\mu\mu}$ 
will allow an unambiguous extraction of the phase $2\varphi_P$, which in turn, with 
the help of $\overline{R}$, will provide the value of $|P|$.

The $P$ parameter can also conveniently be expressed as $P=1+\tilde{P}$ with 
\begin{equation}
	\tilde{P}=|\tilde{P}|e^{i\tilde{\varphi}_P} \equiv
	\frac{\delta C_{10}- C'_{10}}{C_{10}^{\rm SM}} + \frac{m_{B_s}^2}{2m_\mu}\left(\frac{m_b}{m_b + m_s}\right) \left( \frac{C_{P} - C'_{P}}{C_{10}^{\rm SM}}\right).
    \label{PTilde}
\end{equation}
where
\begin{equation}
    \delta C_{10}\equiv C_{10}-C_{10}^{\rm SM}.
\end{equation}
In this notation all NP effects are contained in the parameter $\tilde{P}$.
In the left panel of Figure~\ref{fig:R-ADG_AB} we show the correlations between $\overline{R}$ and ${\cal A}^{\mu\mu}_{\Delta\Gamma}$ in {\it Scenario A} using this notation.
We have varied $\tilde P \in [0,1]$, and most importantly show the strong dependence on the phase $\tilde\varphi_P$.
As will be discussed in detail in Section~\ref{Trees}, the requirement for new gauge bosons or pseudoscalars to satisfy the $B_s$ mixing constraints implies that $\tilde\varphi_P \sim \pi/2$ or $\tilde\varphi_P \sim 0,\pi$, respectively.

Note that in the case of no new phases, $\varphi_P \in \{0, \pi\}$,  and $\phi_s^{\rm NP}=0$,
\be\label{FA3}
{\cal A}^{\mu\mu}_{\Delta\Gamma}=1,\qquad {\cal S}_{\mu\mu}=0, 
\qquad \overline{R}=|P|^2.
\ee
While the first two results coincide with the SM, NP effects can still arise in $\overline{R}$.

\subsubsection{Examples of Models}

\subsubsection*{Constrained Minimal Flavour Violation (CMFV)}

In the CMFV scenario it is assumed that new low-energy effective operators beyond those present in the SM are very strongly suppressed and that flavour violation and CP-violation are governed by the CKM matrix \cite{Buras:2000dm,Buras:2003jf}.
Thus all the Wilson coefficients aside from $C_{10}$ are zero, and $C_{10}$ is real.
This translates into {\it Scenario A}, with the added restrictions that 
$\varphi_P=\phi_s^{\rm NP}=0$. Consequently the formula (\ref{FA3}) applies 
and NP enters only through the ratio $\overline{R}$.

\subsubsection*{Littlest Higgs Model with T-Parity (LHT)}
Similar to CMFV, only SM operators are relevant in this framework but due 
to the presence of new phases in the interactions of SM quarks with mirror 
quarks, CP asymmetries can differ from the SM ones. Therefore  general 
formulae (\ref{FA1}) and (\ref{FA2}) apply here.
Typically ${\rm BR}(B_s\to\mu^+\mu^-)$ is predicted to be larger than its SM 
value but it can only be enhanced by $30\%$ at most
\cite{Blanke:2009am}. A significant part of this enhancement comes 
from the T-even sector that corresponds to the CMFV part of this model, while $|S_{\psi\phi}|$ and $|{\mathcal S}_{\mu\mu}|$, governed by new  phases in the mirror 
quark sector, are at most $0.2$.
 
\subsubsection*{$\boldsymbol{Z'}$ Models and RSc}

As demonstrated in Ref.~\cite{Buras:2012jb}, larger effects than in LHT 
can be found in 
$Z'$ models with tree-level FCNC couplings. If 
new heavy neutral gauge bosons dominate NP contributions to FCNCs, $S=0$ in 
these models,
placing them automatically in {\it Scenario A}. In contrast to 
CMFV and LHT, the presence of new operators implies a rather rich 
phenomenology. Yet, in the absence of $S$
the time-dependent observables are only sensitive to new CP violating phases and are not independent of one another.
In Ref.~\cite{Buras:2012jb} correlations between the ${\cal S}_{\mu\mu}$ observable and $\Delta F=2$ observables have been found for different scenarios for 
$Z'$ couplings with the size of effects that could be measured in the future
provided the masses of these new gauge bosons do not exceed $2$-$3$~TeV. We 
will return to this scenario in Section~\ref{Trees} showing results 
complementary to the ones presented in Ref.~\cite{Buras:2012jb}.

Smaller, but still measurable, effects  have been found in
331 models in which new CP phases are present but no new 
operators~\cite{Buras:2012dp}. Here NP effects in $B_s\to\mu^+\mu^-$ are comparable to the 
ones in the LHT 
model provided  the mass of the new neutral 
gauge boson does not exceed $2$~TeV. 

Finally we mention the Randall--Sundrum model with custodial 
protection in which NP contributions to $B_s\to\mu^+\mu^-$ are governed 
by right-handed flavour-violating $Z$ couplings to quarks but the resulting branching 
ratio is SM-like, with departures from SM prediction at most of order $15\%$
\cite{Blanke:2008yr}. Larger effects are found if the custodial protection 
is absent and then left-handed couplings dominate \cite{Bauer:2009cf}. 
Recently detailed analyses of $Z$ couplings in similar scenarios 
related to partial compositeness have been presented in Refs.~\cite{Barbieri:2012tu,Straub:2013zca,Guadagnoli:2013mru}.
While having different goals than in Ref.~\cite{Buras:2012jb}, they also demonstrate the power 
of $B_s\to\mu^+\mu^-$ in distinguishing between various NP scenarios.

\subsubsection*{Four Generation Models}

In spite of the fact that the existence of a fourth generation seems to be 
very unlikely in view of the LHC data, in particular Higgs branching ratios, 
we just mention that it also belongs to {\it Scenario A}. NP effects in $B_s\to\mu^+\mu^-$ can still be sizable in these models. See Ref.~\cite{Buras:2010pi} and references therein.

\subsubsection*{Pseudoscalar Dominance}

Also a model with NP dominated by tree-level FCNC contributions of 
a pseudoscalar 
belongs to this class. It has been analysed recently in Ref.~\cite{Buras:2013rqa} 
and we will present complementary implications of this model particularly 
suited to our paper in Section~\ref{Trees}.

\subsection{Scenario B: $\boldsymbol{P= 1}$}
\subsubsection{General Formulation}

The simplest realisation of this scenario is $C_{10}=C_{10}^{\rm SM}$ and $C'_{10}=C^{(\prime)}_P=0$.
However, pseudoscalars that couple left-right symmetrically to quarks, so that $C_P=C'_P$, or a conspiracy of the form $C_{10}-C'_{10}=C_{10}^{\rm SM}$ are also allowed.
The point is that in this scenario only scalar operators ${\cal O}^{(\prime)}_S$ drive new physics effects in $\Bsmumu$. In this sense this case is complementary  to {\it Scenario A}.

As there are scalar operators present, there is a rate asymmetry in the $B^0_s$ and $\bar B^0_s$ decays to the individual muon helicities.
Therefore the two time-dependent observables do carry independent information.
In this scenario the observables are given by
\begin{align}\label{SB}
	{\cal A}^{\mu\mu}_{\Delta\Gamma} &= \frac{\cos\phi_s^{\rm NP} - |S|^2\cos(2\varphi_S - \phi_s^{\rm NP})}{1+|S|^2},\notag\\
	{\cal S}_{\mu\mu} &= \frac{-\sin\phi_s^{\rm NP} - |S|^2\sin(2\varphi_S - \phi_s^{\rm NP})}{1+|S|^2},\notag\\
	\overline{R} &= \frac{1+y_s\cos\phi_s^{\rm NP}}{1+y_s} + |S|^2 \left[\frac{1-y_s\cos(2\varphi_S - \phi_s^{\rm NP})}{1+y_s}\right].
\end{align}
Again, with precise value of $\phi_s^{\rm NP}$ to be determined first, these three observables are in principle sufficient to determine the two NP unknowns, $2\varphi_S$ and $|S|$. 
Consequently the untagged observables $\overline{R}$ and ${\cal A}^{\mu\mu}_{\Delta\Gamma}$ are already sufficient to determine  $2\varphi_S$ and $|S|$. 
Moreover, if all three observables are considered, correlations between them 
will result that depend on the precise value of  $\phi_s^{\rm NP}$ \cite{Buras:2013rqa}. 

\begin{figure}[t]
\begin{center}
\includegraphics[height=6.0cm]{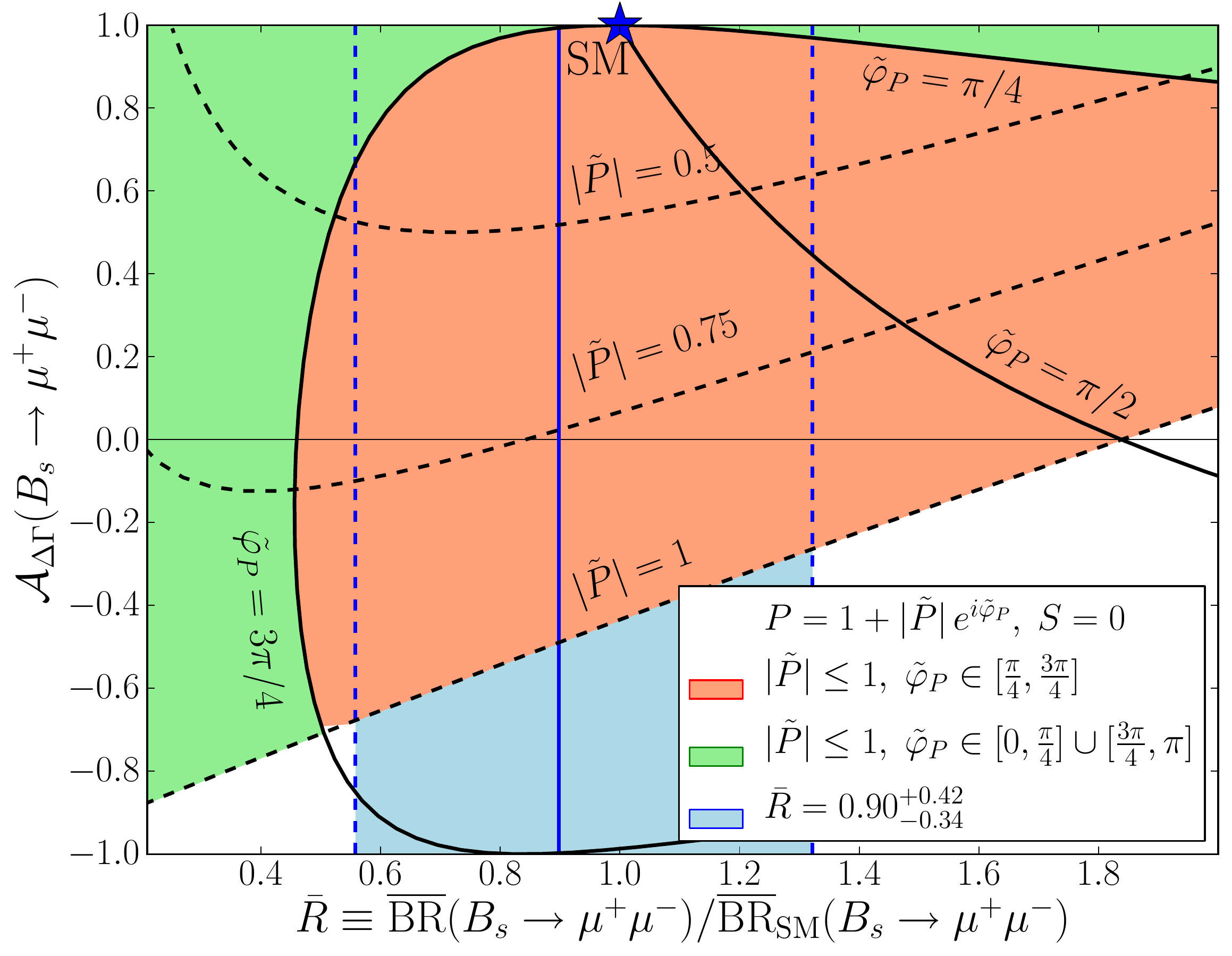}
\includegraphics[height=6.0cm]{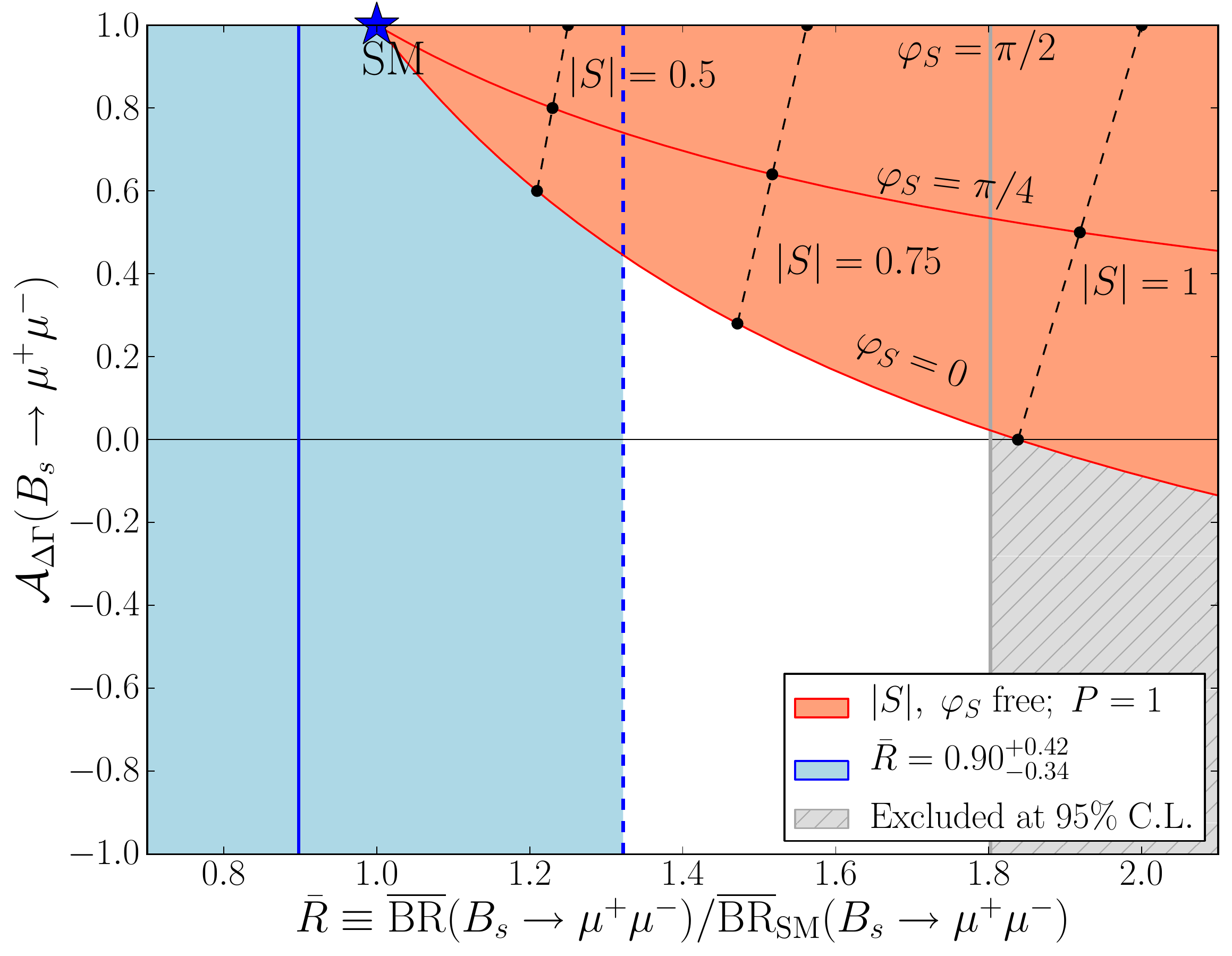}
\end{center}
\caption{The correlation between the $\overline{R}$ and ${\cal A}^{\mu\mu}_{\Delta\Gamma}$ observables in {\it Scenario A} (left panel) and {\it Scenario B} (right panel). 
In {\it Scenario A} we have set $P = 1 + \tilde P$ and $S=0$ with $\tilde P$ free to vary. In {\it Scenario B} $P=1$ and $S$ is free to vary.}
\label{fig:R-ADG_AB}
\end{figure}

In the right panel of Figure~\ref{fig:R-ADG_AB} we show the correlation between $\overline{R}$ and ${\cal A}^{\mu\mu}_{\Delta\Gamma}$ for different values of $S$~\cite{deBruyn:2012wk}.
An interesting feature is that for no CP violating phase, $\varphi_S=\{0,\pi\}$, an increase of $|S|$ pushes ${\cal A}^{\mu\mu}_{\Delta\Gamma}\to 0$.
But within current experimental bounds we have the prediction that ${\cal A}^{\mu\mu}_{\Delta\Gamma}$ cannot take a negative value. Moreover in this scenario 
$|S|\le 0.5$ is favoured.

\subsubsection{Examples of Models}
\subsubsection*{Scalar Dominance}

A  model with NP dominated by tree-level FCNC contributions of 
a scalar belongs to this class. It has been analysed recently in Ref.~\cite{Buras:2013rqa} 
and we will present complementary implications of this model particularly 
suited to our paper in Section~\ref{Trees}.

\subsection{Scenario C: $\boldsymbol{P\pm S= 1}$\label{sec:ScenC}}
\subsubsection{General Formulation}

The meaning of this scenario is clearer if we let $P=1+\tilde{P}$ , with $\tilde P$ defined in \eqref{PTilde}. Then the condition $P\pm S=
1$ is equivalent to $\tilde{P}=\mp S$ i.e.\ in this scenario NP effects to $S$ and $P$ are on the same footing.
If we neglect contributions to $C^{(\prime)}_{10}$ and $m_\mu$ with respect 
to $m_{B_s}$, this scenario is realised if $C^{(\prime)}_S=\pm C^{(\prime)}_P$.

Letting $\tilde{P}=- \kappa S$ for $\kappa=\pm 1$, the time dependent observables are
\begin{align}
	{\cal A}^{\mu\mu}_{\Delta\Gamma} &= \frac{\cos\phi_s^{\rm NP} - 2\,\kappa|S|\cos(\varphi_S - \phi_s^{\rm NP})}
		{1- 2\kappa|S|\cos\varphi_S +2|S|^2 },\notag\\
	{\cal S}_{\mu\mu} &= \frac{-\sin\phi_s^{\rm NP} - 2\,\kappa|S|\sin(\varphi_S - \phi_s^{\rm NP})}
		{1- 2\kappa|S|\cos\varphi_S+2|S|^2 },
\end{align}
which are in general independent.
The branching ratio observable is
\begin{equation}
	\overline{R} = \frac{1-2 \kappa|S|\cos\varphi_S + 2|S|^2 + y_s [\cos\phi_s^{\rm NP}-
	2\kappa|S|\cos(\varphi_S-\phi_s^{\rm NP})]}{1+y_s}.
\end{equation}
In the presence of a precise value of $\phi_s^{\rm NP}$, these three observables are sufficient to determine the two NP unknowns $\varphi_S$ and $|S|$. Moreover,
correlations between the involved observables characteristic for this scenario 
and additional tests are possible.

The observable $\overline{R}$ is minimised by $S_{\rm crit}= \kappa(1+y_s)/2$ and $\phi_s^{\rm NP}=0$, giving the lower bound
\begin{equation}
	\overline{R} \geq \frac{1-y_s}{2}.
	\label{RlowerBound}
\end{equation}
This lower bound, without the $y_s$ and phase considerations, was first observed in Ref.~\cite{Logan:2000iv}.
A branching ratio measurement below this bound would thereby rule out this scenario.

If we assume the new physics phase $\phi_s^{\rm NP}$ in $B_s$ mixing is known, then the purely untagged observables ${\cal A}^{\mu\mu}_{\Delta\Gamma}$ and $\overline{R}$ can solve for $S$ and $\varphi_S$.
Setting $\phi_s^{\rm NP}=0$ for simplicity, we have the expressions
\begin{align}
	|S| = |P-1| &= \sqrt{\frac{\overline{R}\,(1+y_s)(1-{\cal A}^{\mu\mu}_{\Delta\Gamma})}{2(1+y_s\,{\cal A}^{\mu\mu}_{\Delta\Gamma})}},\notag\\
	\cos\varphi_S = -\kappa\cos(\tilde{\varphi}_P)&= \sqrt{\frac{(1+y_s\,{\cal A}^{\mu\mu}_{\Delta\Gamma})}{2\,\overline{R}\,(1+y_s)(1-{\cal A}^{\mu\mu}_{\Delta\Gamma})}}\left[1- \frac{\overline{R}\,(1+y_s){\cal A}^{\mu\mu}_{\Delta\Gamma}}{1+y_s{\cal A}^{\mu\mu}_{\Delta\Gamma}} \right].
\end{align}

\begin{figure}[t]
\begin{center}
\includegraphics[height=6.0cm]{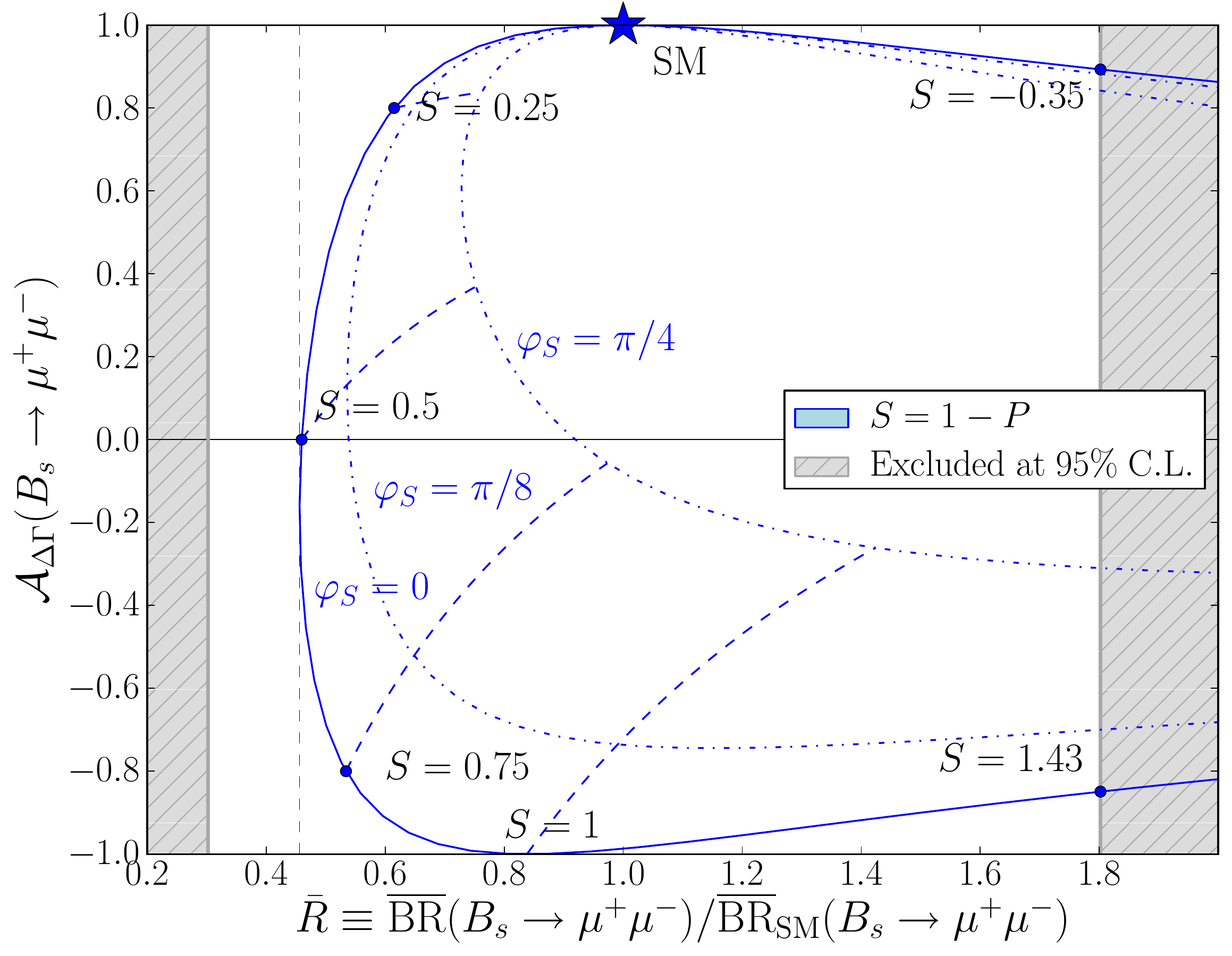}
\includegraphics[height=6.0cm]{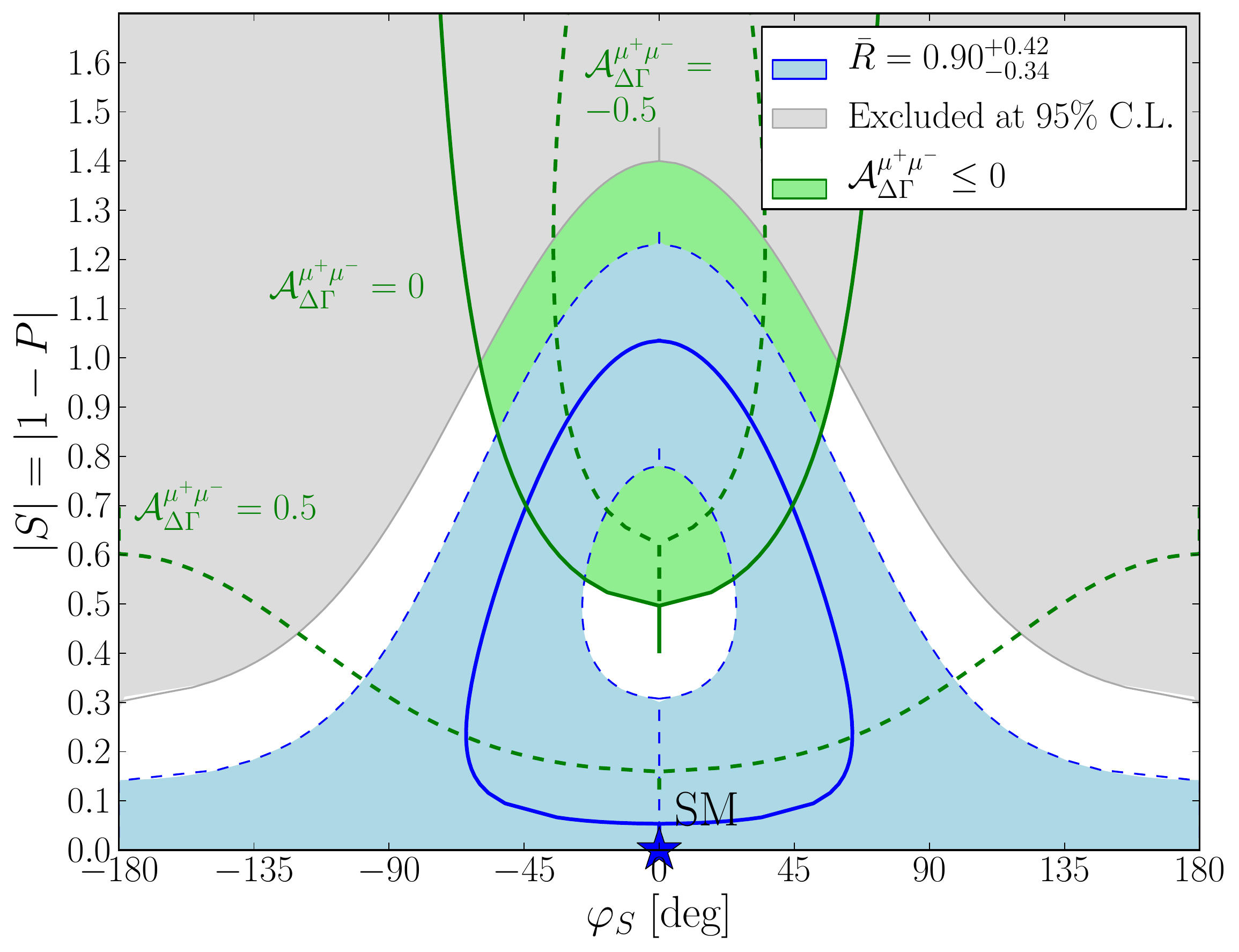}
\end{center}
\caption{Scenario C: $P\pm S =1$. {\it Left panel:} the correlation between the $\overline{R}$ and ${\cal A}^{\mu\mu}_{\Delta\Gamma}$ observables. {\it Right panel:} correlation between the $|S|=|P-1|$ and $\varphi_S= \tilde{\varphi_P}+ (1+\kappa)\pi/2$ NP parameters (see text).}
\label{fig:SnP}
\end{figure}

In the left panel of Figure~\ref{fig:SnP} we show the correlation between $\overline{R}$ and ${\cal A}^{\mu\mu}_{\Delta\Gamma}$ in the limit $\phi_s^{\rm NP}=0$.
Observe the lower bound on $\overline{R}$ specified in \eqref{RlowerBound}.
If, furthermore, $\tilde{\varphi_P}=\varphi_S=\{0,\pi\}$ we observe that ${\cal A}^{\mu\mu}_{\Delta\Gamma}$ can help to resolve the two possible solutions for $S$ coming from a branching ratio measurement $\overline{R}$.

In the right panel of Figure~\ref{fig:SnP} we show the correlation between $\varphi_S$ and $|S|$.
Observe that the current measurement of $\overline{R}$ still allows a large range for both NP parameters.
If ${\cal A}^{\mu\mu}_{\Delta\Gamma}$ were measured with a negative sign it would indicate large contributions from NP.
Moreover in this case the ${\cal A}^{\mu\mu}_{\Delta\Gamma}$ sharply cuts the $\overline{R}$ contour, so that a measurement of ${\cal
A}^{\mu\mu}_{\Delta\Gamma}$ would distinguish between the magnitude and the phase of $S$ up to the twofold ambiguity in
$\varphi_S$.

\subsubsection{Examples of Models}
\subsubsection*{Two Higgs-Doublet Models (2HDM), MSSM}
A 2HDM in the decoupling regime, such that $M_{H^0}\simeq M_{A^0}\simeq M_{H^\pm}\gg M_h$~\cite{Gunion:2002zf}, has the generic feature that
\begin{equation}\label{2HDM}
	C_S = -C_P,\quad C'_S=C'_P.
\end{equation}
If the couplings of the heavy Higgs bosons are not left-right symmetric, so that either $C_{S,P}$ or $C'_{S,P}$ are dominant\footnote{In MFV this is the case. Namely $C'_{S,P}/C_{S,P}\sim m_s/m_b$.}, this corresponds to {\it Scenario C}.
Thus the branching ratio has a lower bound and a significant scalar NP contribution is indicated by negative values of ${\cal A}^{\mu\mu}_{\Delta\Gamma}$. A precise measurement of the untagged observable ${\cal A}^{\mu\mu}_{\Delta\Gamma}$ can distinguish the phase and magnitude of the NP Wilson coefficients. 
We will analyse a similar scenario in more detail in Section~\ref{Trees}.

The above is true also for the MSSM, provided that NP contributions to vector-axial operators, $C'_{10}$, are negligible.
The MSSM has the added advantage that large $\tan\beta$ effects, which are one way to realise the decoupling regime, can give a significant boost to the scalar operators~\cite{Babu:1999hn,Isidori:2001fv,Buras:2002vd}. 

If the 2HDM is not in a decoupling regime, then either the physical scalar $H^0$ or pseudoscalar $A^0$ may be considerably lighter than the other.
If this solo particle can generate the required FCNC, then we are in {\it Scenario B} or {\it Scenario A} respectively.

\subsection{Scenario D: $\boldsymbol{\varphi_P,\varphi_S\in \{0,\pi\}}$}
\subsubsection{General Formulation}
In this scenario we assume no CP violating phases in the $\Bsmumu$ decay mode: $\varphi_P,\varphi_S\in \{0,\pi\}$ \cite{deBruyn:2012wk}.
This is equivalent to all of the Wilson coefficients taking real values.
Clearly this constraint can also be applied to the other scenarios discussed in this section, but this scenario is distinct in that $S$ and $P$ are allowed to remain arbitrary {\it real} values. 
Yet, in the presence of a non-vanishing  NP phase $\phi_s^{\rm NP}$, the CP-asymmetry ${\cal S}_{\mu\mu}$ could be non-vanishing.

The resulting time dependent observables in this scenario are
\begin{align}
	{\cal A}^{\mu\mu}_{\Delta\Gamma} = \cos\phi_s^{\rm NP}\left[
	\frac{|P|^2 -|S|^2}{|P|^2+|S|^2}\right],\quad
	{\cal S}_{\mu\mu} = -\sin\phi_s^{\rm NP}\left[\frac{|P|^2 -|S|^2}{|P|^2+|S|^2}\right],
\end{align}
and the branching ratio observable is given by
\begin{align}
	\overline{R} = |P|^2\left[\frac{1+y_s\cos\phi_s^{\rm NP}}{1+y_s}\right] + |S|^2 \left[\frac{1-y_s\cos\phi_s^{\rm NP}}{1+y_s}\right].
\end{align}

Importantly, whereas the branching ratio observable $\overline{R}$ gives their squared sum, the ${\cal A}^{\mu\mu}_{\Delta\Gamma}$ is sensitive to the difference.
With known $\phi_s^{\rm NP}$
these three observables are sufficient to determine the two NP unknowns 
 $|P|$ and $|S|$. As $\sin\phi_s^{\rm NP}$ is already known to be 
small, ${\cal S}_{\mu\mu}$ is also small in this scenario. Consequently ${\cal A}^{\mu\mu}_{\Delta\Gamma}$ and $\overline{R}$ will 
be the relevant observables in this determination.
With  $\cos\phi_s^{\rm NP}$ very close to 
unity one finds then
\be\label{PSResult}
|P|^2=(1+y_s)\frac{\overline{R}}{2}\left[\frac{1+{\cal A}^{\mu\mu}_{\Delta\Gamma}}{1+y_s{\cal A}^{\mu\mu}_{\Delta\Gamma}}\right], \qquad 
|S|^2=(1+y_s)\frac{\overline{R}}{2}\left[\frac{1-{\cal A}^{\mu\mu}_{\Delta\Gamma}}{1+y_s{\cal A}^{\mu\mu}_{\Delta\Gamma}}\right].
\ee
Finally a measurement of ${\cal S}_{\mu\mu}$ incompatible with the known value of $\phi_s^{\rm NP}$ would exclude this scenario and indicate new CP violating phases in the decay.

\begin{figure}[t]
\begin{center}
\includegraphics[height=6.0cm]{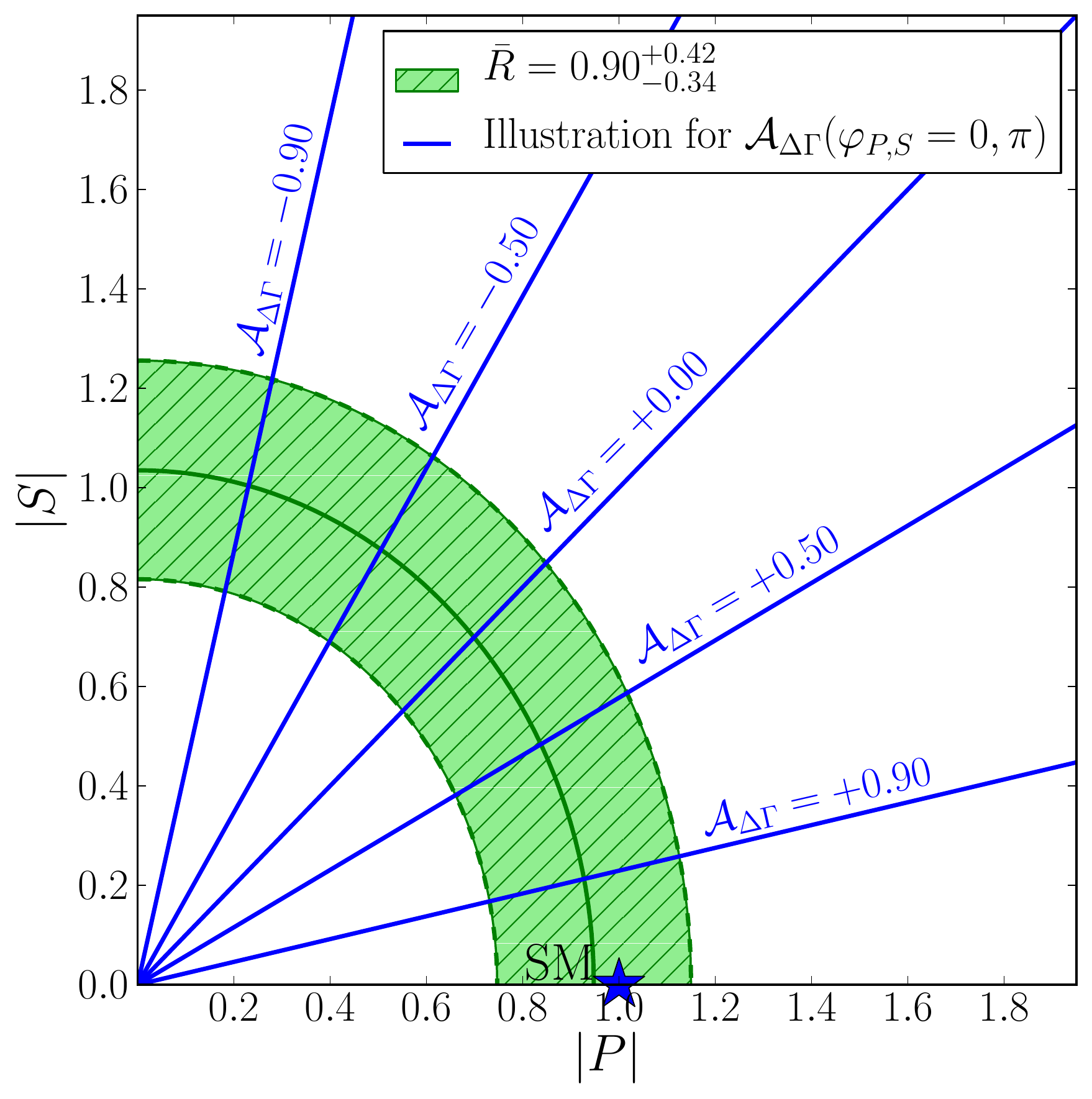}
\end{center}
\caption{Scenario D: $\varphi_P,\varphi_S\in \{0,\pi\}$. The correlation between the $|P|$ and $|S|$ parameters for varying values of ${\cal A}^{\mu\mu}_{\Delta\Gamma}$. Also shown is the current measurement of $\overline{R}$.}
\label{fig:scenD}
\end{figure}

In Figure~\ref{fig:scenD} we illustrate how measurements of $\overline{R}$ and ${\cal A}^{\mu\mu}_{\Delta\Gamma}$ can be used to pinpoint the parameters $|S|$ and $|P|$ (we have taken $\phi_s^{\rm NP}=0$).

\subsubsection{Example of Models}

\subsubsection*{Minimal Flavour Violation (MFV)}

Models with Minimal Flavour Violation (MFV), but without flavour 
blind phases, as formulated as an effective field theory in Ref.~\cite{D'Ambrosio:2002ex}, belong naturally to this class.
MFV protects against any additional flavour structure or CP violation beyond what is already present in the CKM matrix, while still allowing for additional, higher-dimensional, operators~\cite{D'Ambrosio:2002ex}.
MFV therefore falls into {\it Scenario D}, with the added restriction that also $\phi_s^{\rm NP}$ is zero.
Thus in models with MFV, as seen in (\ref{PSResult}), the time-dependent untagged observable ${\cal A}^{\mu\mu}_{\Delta\Gamma}$ together with the branching ratio observable $\overline{R}$ are sufficient to disentangle the scalar contribution $S$ from $P$.
A measurement of ${\cal S}_{\mu\mu}\neq 0$ would falsify MFV. 
Typical examples in this class are MSSM with MFV and 2HDM with MFV.

An exception are models with  MFV and flavour-blind phases, like the 
2HDM with such phases, also known as ${\rm 2HDM_{\overline{\rm MFV}}}$~\cite{Buras:2010mh}. 
In this case model specific details are necessary in order for the time-dependent observables to distinguish between the operators and phases.

\subsection{Scenario E: $\boldsymbol{P= 0}$}

In this scenario $C^{(\prime)}_P$, $C'_{10}$ or $\delta C_{10}$ destructively interfere with $C_{10}^{SM}$ to drive $P$ to zero.
Then non-zero values of the $\Bsmumu$ observables will be driven purely by the operators ${\cal O}^{(\prime)}_S$. 

This scenario is similar to {\it Scenario A}, in that there is no rate asymmetry between the individual helicity decay modes. Thus the time-dependent observables are not independent:
\begin{align}
	{\cal A}^{\mu\mu}_{\Delta\Gamma} = -\cos(2\varphi_S - \phi_s^{\rm NP}),\quad
	{\cal S}_{\mu\mu} = -\sin(2\varphi_S - \phi_s^{\rm NP}).
\end{align}
The key difference, however, is that now only a scalar and not a gauge boson or pseudoscalar is 
at work. Moreover,  in the absence of new CP-violating phases 
${\cal A}^{\mu\mu}_{\Delta\Gamma}=-1$, which differs by sign from the Standard Model value and the analogous case in {\it Scenario A} as seen in (\ref{FA3}).
This is also a limiting case of {\it Scenario D}.
The branching ratio observable is given by
\begin{equation}
	\overline{R} = |S|^2 \left[\frac{1-y_s\cos(2\varphi_S - \phi_s^{\rm NP})}{1+y_s}\right].
\end{equation}

We do not know any specific model that would naturally be placed in this 
scenario but we will investigate in Section~\ref{Trees} whether 
requiring tree-level exchanges of a pseudoscalar to cancel SM contribution
is still 
consistent with the data.

\subsection{Summary}
In Table~\ref{tab:Models} we collect the properties of the selected models discussed above with 
respect to the basic phenomenological parameters listed in (\ref{unknowns})
and the class they belong to. We also indicate whether the phase $\phi^{\rm NP}_s$ can be non-zero in these models.
In 
all cases $|P|$ is generally different from zero as it contains the SM contributions. In order to distinguish between different models in each row of this table a more detailed analysis has to be performed taking all existing constraints 
into account. However, already identifying which of these four rows has been chosen by nature would be a tremendous step forward.

\begin{table}[t]
\centering
\begin{tabular}{|c||c||c|c|c|c|c|}
\hline
 Model & Scenario & $|P|$ & $\varphi_P$ & $|S|$ &  $\varphi_S$   & $\phi_s^{\rm NP}$\\
\hline
\hline
  \parbox[0pt][1.6em][c]{0cm}{} CMFV & A & $|P|$  &  $0$ & $0$ & $0$  & $0$ \\
 \parbox[0pt][1.6em][c]{0cm}{} MFV & D & $|P|$  &  $0$ & $|S|$ & $0$  & $0$ \\
 \parbox[0pt][1.6em][c]{0cm}{} LHT,~4G,~RSc,~$Z'$ & A & $|P|$  &  $\varphi_P$ & $0$ & $0$  & $\phi_s^{\rm NP}$ \\
 \parbox[0pt][1.6em][c]{0cm}{} 2HDM (Decoupling) & C & $|1\mp S|$  &   ${\rm arg}(1\mp S)$ & $|S|$
 & $\varphi_S$  & $\phi_s^{\rm NP}$ \\
\parbox[0pt][1.6em][c]{0cm}{} 2HDM (A Dominance) & A & $|P|$  &   $\varphi_P$ & $0$
 & $0$  & $\phi_s^{\rm NP}$ \\
\parbox[0pt][1.6em][c]{0cm}{} 2HDM (H Dominance) & B & $1$  &   $0$ & $|S|$
 & $\varphi_S$  & $\phi_s^{\rm NP}$ \\
\hline
\end{tabular}
\caption{General structure of basic variables in different NP models. The last three cases apply also to the MSSM.}
\label{tab:Models}
\end{table}

\section{Specific Models and Constraints from $\boldsymbol{B_s}$ Mixing}\label{Trees}
\subsection{Tree-Level Neutral Gauge Boson Exchange}
\subsubsection{Basic Formulae}
As the first class of specific models we consider $Z'$ models in which 
NP contributions to FCNC observables are dominated by tree-level $Z'$ 
exchanges. A detailed analysis of these models has recently been 
presented in Ref.~\cite{Buras:2012jb}. Also there the three observables in 
(\ref{trio}) have been considered but the emphasis has been put on 
the correlations of them with $\Delta F=2$ observables, in particular 
$S_{\psi\phi}$. Here we will complement this study by computing the correlations 
among $\overline{R}$, $\mathcal{A}^{\mu\mu}_{\Delta\Gamma}$, 
and ${\cal S}_{\mu\mu}$, while taking the constraints from $\Delta F=2$ 
observables obtained in Ref.~\cite{Buras:2012jb} into account.

We define the flavour-violating couplings of $Z'$ 
to quarks as follows
\be\label{eq:3.14}
 \mathcal{L}_\text{FCNC}(Z')=\left[\Delta_L^{sb}(Z')(\bar s \gamma_\mu P_L b)+
                      \Delta_R^{sb}(Z')(\bar s \gamma_\mu P_R b)\right] Z^{'\mu},
\ee
where $\Delta_{L,R}^{sb}(Z')$ are generally complex.

We also define the $Z'$ couplings to muons 
\begin{equation}
\mathcal{L}_{\ell \bar \ell}(Z')= \left[\Delta_L^{\ell\ell}(Z')
(\bar\ell\gamma_\mu P_L\ell)
+\Delta_R^{\ell\ell}(Z')(\bar\ell\gamma_\mu P_R\ell)\right]
Z^{'\mu}\,
\end{equation}
and introduce 
\be\label{DeltasVA}
\Delta_A^{\mu\bar\mu}(Z')= \Delta_R^{\mu\bar\mu}(Z')-\Delta_L^{\mu\bar\mu}(Z').
\ee

Then the non-vanishing Wilson coefficients contributing to $B_s\to\mu^+\mu^-$ 
are given as follows:
\begin{align}
    \sin^2\theta_W C_{10} &= -\eta_Y Y_0(x_t) -\frac{1}{g_{\text{SM}}^2}\frac{1}{M_{Z'}^2}
\frac{\Delta_L^{sb}(Z^\prime)\Delta_A^{\mu\bar\mu}(Z')}{V_{ts}^* V_{tb}}, \label{C10}\\
   \sin^2\theta_W C_{10}^\prime   &= -\frac{1}{g_{\text{SM}}^2}\frac{1}{M_{Z'}^2}
\frac{\Delta_R^{sb}(Z')\Delta_A^{\mu\bar\mu}(Z')}{V_{ts}^* V_{tb}},\label{C10P}
 \end{align}
where
\be\label{gsm}
g_{\text{SM}}^2=4\frac{G_F}{\sqrt 2}\frac{\alpha}{2\pi\sin^2\theta_W}\,.
\ee

As only the coefficients $C_{10}$ and $C_{10}'$ are non-vanishing this NP 
scenario is governed by the formulae (\ref{FA1}) and (\ref{FA2}). Indeed 
this scenario is an example of {\it Scenario A} in which, in addition to $S=0$, 
also the pseudoscalar contributions vanish. Yet, as $P$ can differ from unity 
and have a nontrivial phase, a rich phenomenology is found \cite{Buras:2012jb}.

\subsubsection{Numerical Analysis}
It is not our goal to present a full-fledged numerical
analysis of all correlations including present theoretical, parametric and experimental
uncertainties as this would only wash out the effects we want to emphasize. Therefore we simply choose the three parameters entering our formulae,
$F_{B_s}$, $\tau(B_s)$ and $|V_{ts}|$ to be in the ballpark of 
their present central values:
 \be
F_{B_s}=225.0\mev, \quad \tau(B_s)=1.503~{\rm ps}, \quad  |V_{ts}|=0.040.
\ee
Other relevant input can be found in the Tables of Ref.~\cite{Buras:2013rqa}.

The main theoretical uncertainties in our analysis are due to the constraints 
on the couplings $\Delta_{L,R}^{sb}(Z)$ and $\Delta_{L,R}^{sb}(H)$ coming 
from the experimental values of $\Delta M_s$ and $S_{\psi\phi}$. Indeed the 
hadronic matrix elements of new operators are still subject to significant 
uncertainties. We will not recall the relevant formulae as they can be 
found in Ref.~\cite{Buras:2012jb}.  We will use 
the full machinery presented in that paper, setting the relevant 
parameters at their central values  and requiring $\Delta M_s$ and 
$S_{\psi\phi}$ to be in the ranges
\be\label{C1}
16.9/{\rm ps}\le \Delta M_s\le 18.7/{\rm ps},
\quad  -0.20\le S_{\psi\phi}\le 0.20.
\ee
Concerning the first range, it effectively takes the hadronic uncertainties 
into account. The second range corresponds to the $2\sigma$ range for 
$\phi_s$ in (\ref{phis}).

As far as the direct lower bound  on  $M_{Z'}$  from collider experiments is concerned, the 
most stringent bounds are provided by CMS experiment \cite{Chatrchyan:2012it} but 
these constraints are mainly sensitive  to the couplings of the $Z'$ to the light quarks which 
do not play any role in our analysis. Moreover, the collider bounds on $M_{Z'}$ are 
generally model dependent.  While for the so-called sequential $Z'$ the lower 
bound for $M_{Z'}$ is in the ballpark of $2.5\tev$, in other models values 
as low as $1\tev$ are still possible. In order to cover large set of models, 
we will choose as our nominal value $M_{Z'}=1\tev$. With the help of 
the formulae in Ref.~\cite{Buras:2012jb} it is possible to 
estimate approximately, how our results would change for 
$1\tev\le M_{Z'}\le 3\tev$.

As in the analyses in Refs.~\cite{Buras:2012jb,Buras:2013rqa} it 
will be instructive to consider
 the following four schemes for the gauge boson couplings and  in the next subsection for scalar couplings:
\begin{enumerate}
\item
Left-handed Scheme (LHS) with complex $\Delta_L^{bs}\not=0$  and $\Delta_R^{bs}=0$,
\item
Right-handed Scheme (RHS) with complex $\Delta_R^{bs}\not=0$  and 
$\Delta_L^{bs}=0$,
\item
Left-Right symmetric Scheme (LRS) with complex
$\Delta_L^{bs}=\Delta_R^{bs}\not=0$,
\item
Left-Right asymmetric Scheme (ALRS) with complex
$\Delta_L^{bs}=-\Delta_R^{bs}\not=0$.
\end{enumerate}
Note that the ordering in flavour indices in the couplings in these schemes 
is governed by the operator structure in $B_s^0$--$\bar B_s^0$ mixing 
\cite{Buras:2012jb,Buras:2013rqa} and differs from the one in (\ref{eq:3.14}) 
and (\ref{eq:3.15}). In this context one should recall that
\be\label{recall}
\Delta_{L,R}^{sb}(Z')=[\Delta_{L,R}^{bs}(Z')]^*,\qquad
\Delta_{L,R}^{sb}(H)=[\Delta_{R,L}^{bs}(H)]^*,
\ee
where $H$ stands for either scalar or pseudoscalar.

The ranges for $B_s$ mixing given in \eqref{C1} result in two allowed regions for the magnitudes and phases of the quark couplings
$\Delta^{sb}_{L,R}$ depending on the scheme chosen above.
These regions in parameter space are dubbed {\it oases}. 
The oases for each case have a two fold degeneracy in the complex phase of the coupling.
Where it is relevant we will distinguish between these two different oases using the colours blue and red.

\begin{figure}[!tb]
\centering
\includegraphics[width=0.45\textwidth] {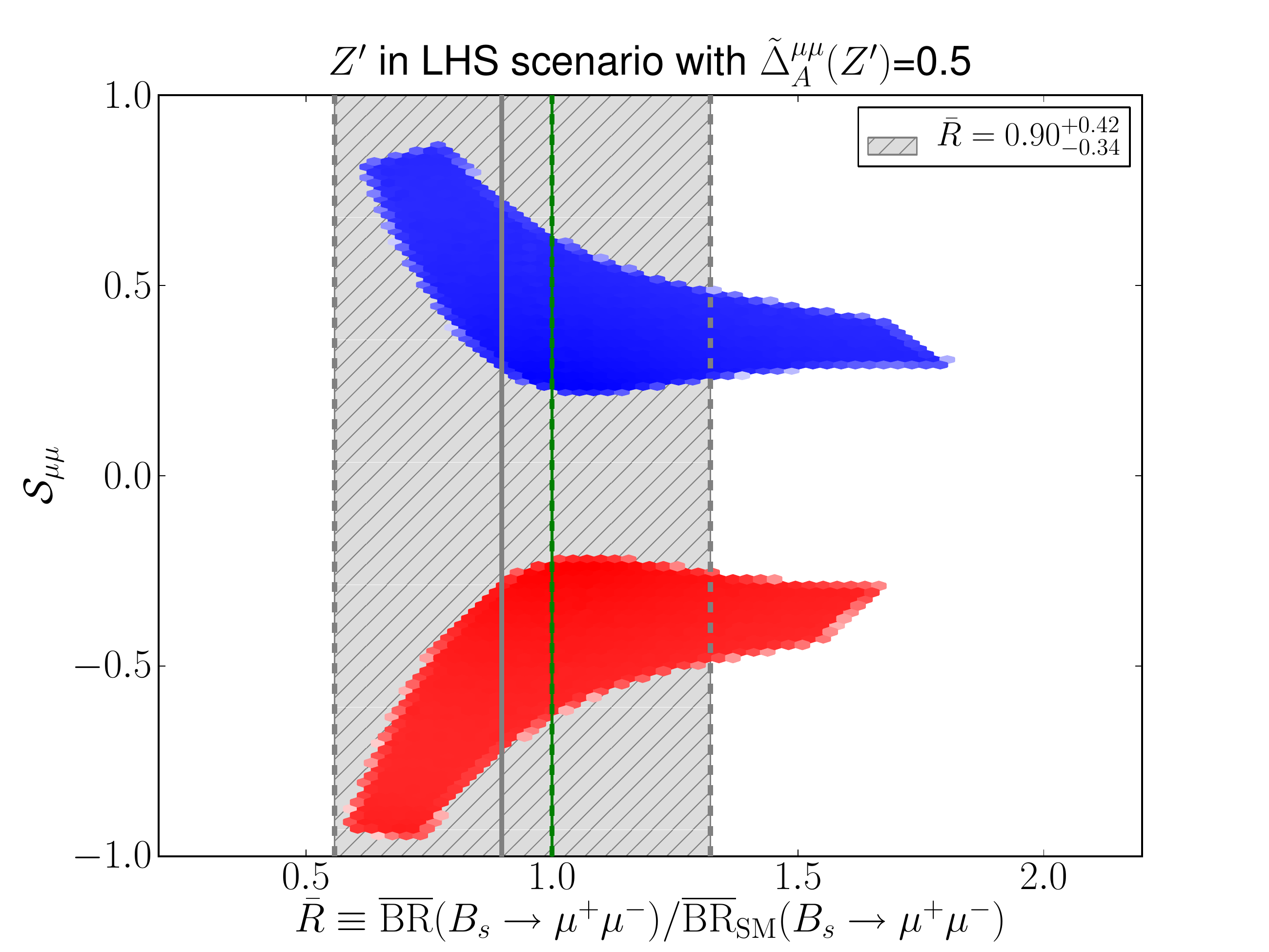}
\includegraphics[width=0.45\textwidth] {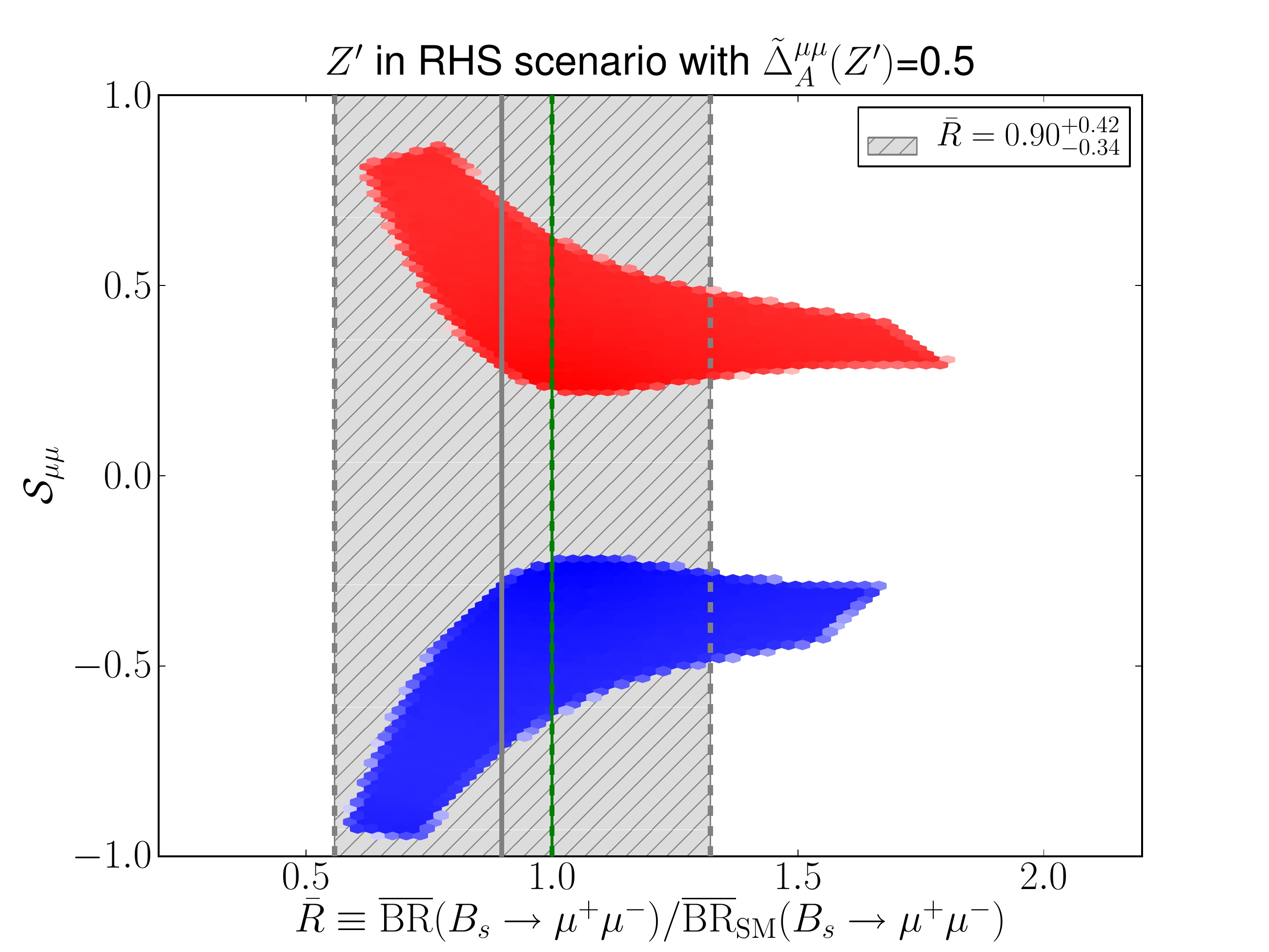}
\caption{${\cal S}_{\mu\mu}$ versus $\overline{R}$
 for LHS (left) and 
 RHS (right), assuming $M_{Z^\prime} = 1~$TeV and $\Delta^{\mu\mu}_A(Z')=0.5$.  
Gray region: exp 1$\sigma$ range  for $\overline{R}$.
The $2\,\sigma$ CL combined fit region for the Wilson coefficients $C^{(\prime)}_{10}$ come from a general $b\to s l^+ l^-$ analysis given in Ref.~\cite{Altmannshofer:2012ir}.
}
\label{fig:Z1}
\end{figure}

\begin{figure}[!tb]
\centering
\includegraphics[width=0.45\textwidth] {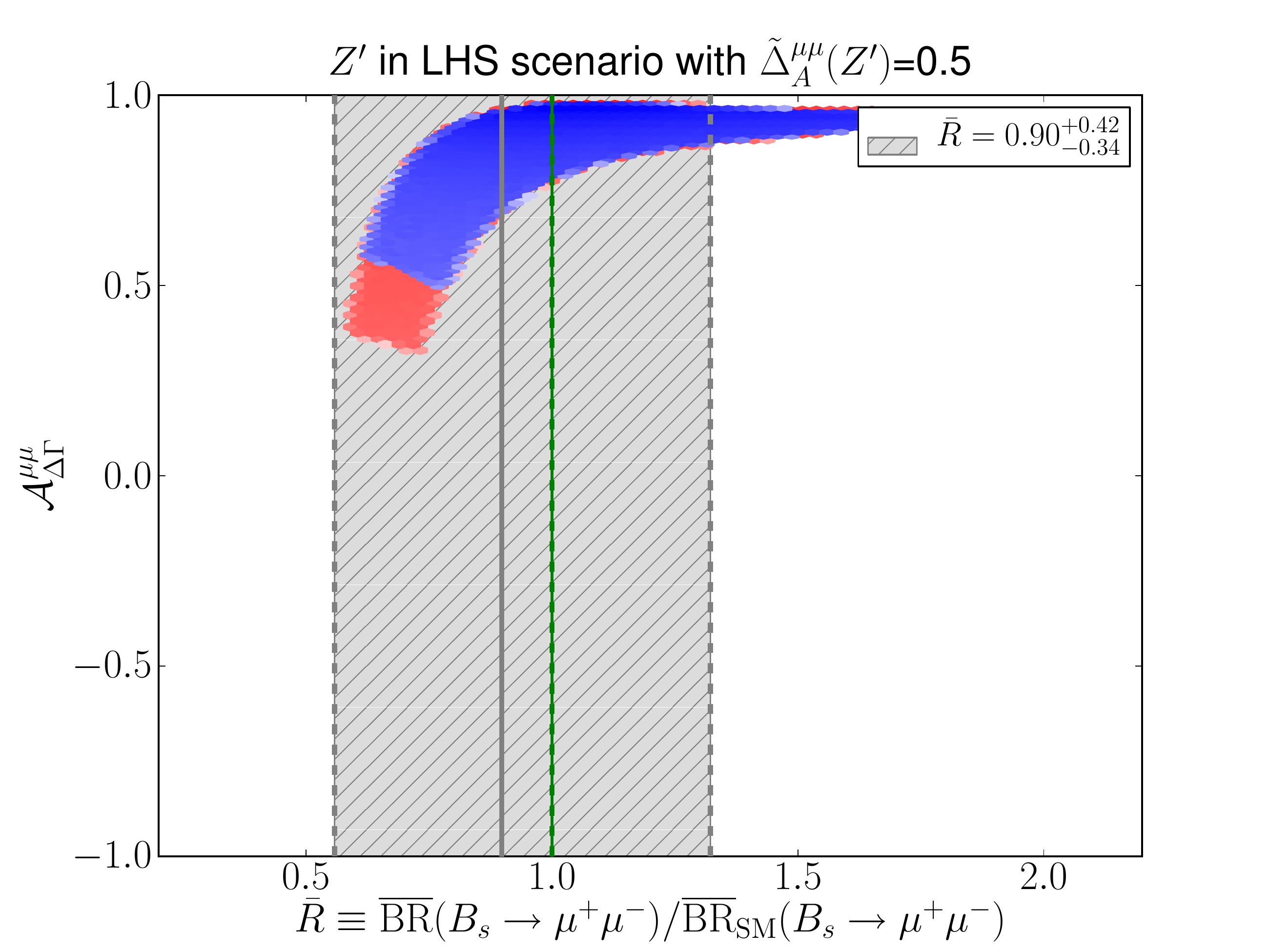}
\includegraphics[width=0.45\textwidth] {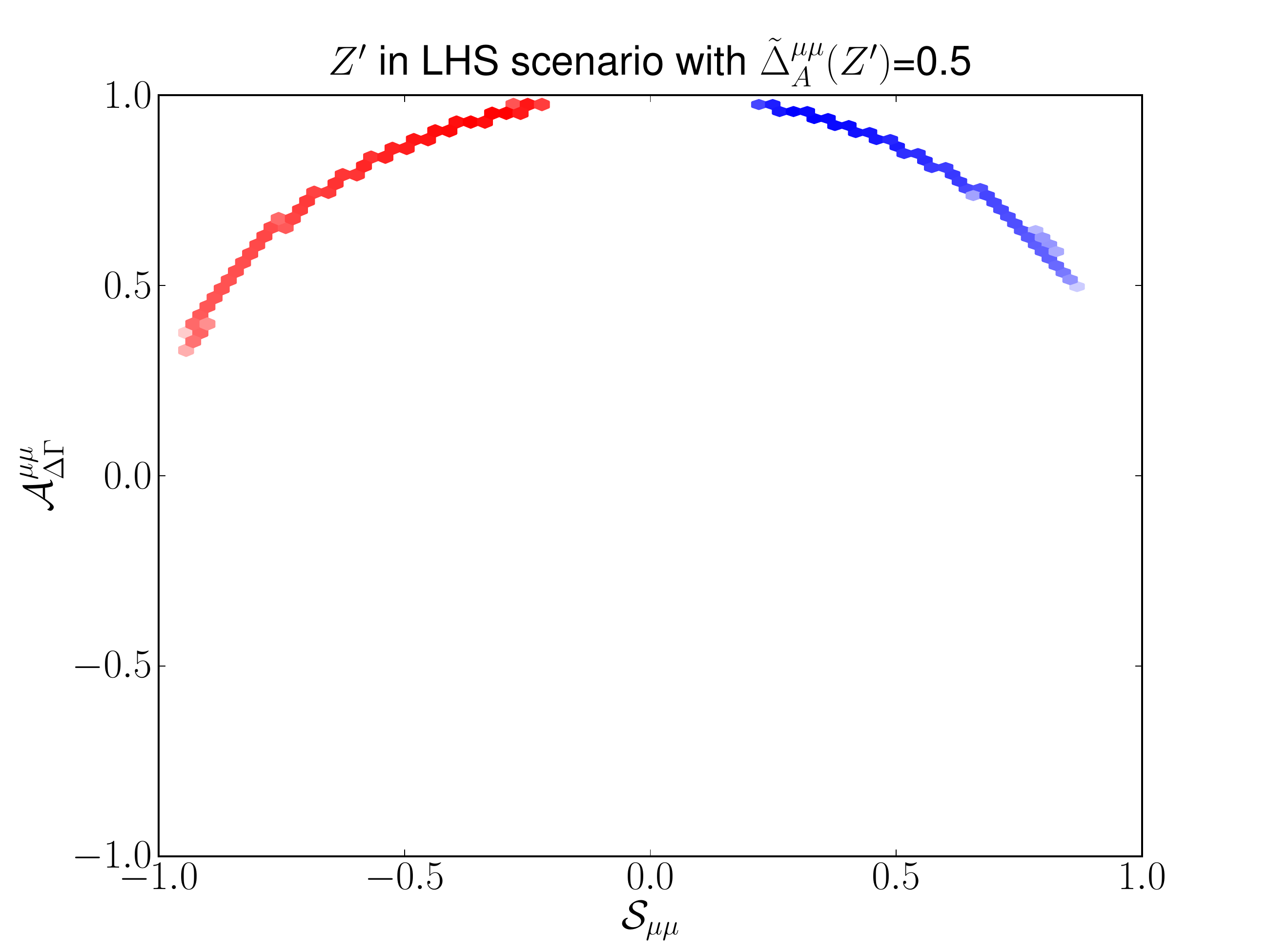}
\caption{$\mathcal{A}^{\mu\mu}_{\Delta\Gamma}$  versus $\overline{R}$ (left) and $\mathcal{A}^{\mu\mu}_{\Delta\Gamma}$ versus ${\cal S}_{\mu\mu}$ (right)
for LHS, assuming $M_{Z^\prime} = 1~$TeV and $\Delta^{\mu\mu}_A(Z')=0.5$.  
Gray region in left panel: exp 1$\sigma$ range  for $\overline{R}$.
The $2\,\sigma$ CL combined fit region for the Wilson coefficient $C_{10}$ comes from a general $b\to s l^+ l^-$ analysis given in Ref.~\cite{Altmannshofer:2012ir}.
}
 \label{fig:Z2}
\end{figure}

In order to perform the present analysis we assign $\Delta_A^{\mu\bar\mu}(Z')=0.5$, as was done in Ref.~\cite{Buras:2012jb}.
In Section~\ref{sec:ZprimeCompare} and beyond, where we compare $Z'$ exchange with various (pseudo)scalar exchanges, this coupling will be allowed to vary.
The  sign of this coupling is crucial for the identification of 
various enhancements
and suppressions with respect to SM branching ratio and CP asymmetries and
impacts the search for successful oases in the space of parameters
that has been performed in Ref.~\cite{Buras:2012jb}.
If the sign of the $Z'$ coupling to muons will be identified in the future to be different from the one
assumed here, it will be straightforward, in combination with the discussion 
in Ref.~\cite{Buras:2012jb},
to find out how our results will be modified. 
In Figure~\ref{fig:Z1} we show the correlation between 
${\cal S}_{\mu\mu}$ and $\overline{R}$ for LHS (left) and RHS (right).
Corresponding correlations between $\mathcal{A}^{\mu\mu}_{\Delta\Gamma}$ and 
$\overline{R}$ and between $\mathcal{A}^{\mu\mu}_{\Delta\Gamma}$ and ${\cal S}_{\mu\mu}$ are given in Figure~\ref{fig:Z2} for LHS only.
The two colours correspond to two oases in the values of the coupling 
$\Delta_{L,R}(Z')$ that are consistent with $\Delta M_s$ and $S_{\psi\phi}$ 
constraints.

We observe in analogy with findings of Ref.~\cite{Buras:2012jb} that 
the correlations in the LHS and RHS schemes have the same shape except
the oases and consequently the colours in Figure~\ref{fig:Z1}
have to be  interchanged. 
We conclude therefore that on the basis of the three observables considered 
by us it is not
possible to distinguish between LHS and RHS schemes because in the
RHS scheme one can simply interchange the oases
 to obtain the same
physical results as in LHS scheme. Consequently 
if one day we will have precise measurements of ${\cal A}^{\mu\mu}_{\Delta\Gamma}$,  $S_{\mu^+\mu^-}$ and ${\rm BR}(B_s\to\mu^+\mu^-)$
we will still not be able to distinguish for instance whether we deal
with LHS scheme in the blue oasis or RHS scheme in the red oasis.

As pointed out in Ref.~\cite{Buras:2012jb}, in order to make this distinction 
one has to consider simultaneously $B\to K^*\mu^+\mu^-$, $B\to K \mu^+\mu^-$ and $b\to s\nu\bar\nu$ transitions, which is beyond the scope of our paper.
However, we do include regions corresponding to the $2\,\sigma$ CL combined fits of Ref.~\cite{Altmannshofer:2012ir} for the Wilson coefficients $C_{10}$ and $C'_{10}$, which result from these transitions, in Figures~\ref{fig:Z1} and \ref{fig:Z2} where relevant.
The combination of our oases and these additional constraints gives us valuable information.
The allowed values for the three observables considered 
are, in this NP scenario,
\be
0.4\le{\cal A}^{\mu\mu}_{\Delta\Gamma}\le 1.0, \quad 0.2\le |{\cal S}_{\mu^+\mu^-}| \le 0.9, \quad 0.5\le \overline{R}\le
\left\{
\begin{array}{ccl}
    1.3 & : & {\rm LHS\ scheme} \\
    1.0 & : & {\rm RHS\ scheme}
\end{array}\right. .
\ee
Moreover, the smallest values of ${\cal A}^{\mu\mu}_{\Delta\Gamma}$ and largest 
values of $|{\cal S}_{\mu^+\mu^-}|$ are obtained for smallest values of $\overline{R}$. 
The non-zero values of ${\cal S}_{\mu^+\mu^-}$ originate in $Z'$ models 
from requiring that $\Delta M_s$ is suppressed with respect to its 
SM value in order to achieve a better agreement with data. As we will see below 
for models with scalar or pseudoscalar exchanges, this requirement can also 
be satisfied for a vanishing ${\cal S}_{\mu^+\mu^-}$.

If both LH and RH currents are present in NP contributions, but
we impose a symmetry between LH and RH quark couplings, then  NP contributions to $B_{s}\to\mu^+\mu^-$ vanish. 
We thus find that 
\begin{align}\label{FB1}
	{\cal A}^{\mu\mu}_{\Delta\Gamma} = \cos(\phi_s^{\rm NP}),\quad
	{\cal S}_{\mu\mu} = -\sin(\phi_s^{\rm NP}).
\end{align}
The branching ratio observable is given by
\begin{equation}\label{FB2}
	\overline{R} = \left[\frac{1+y_s\cos(\phi_s^{\rm NP})}{1+y_s}\right],
\end{equation}
which can also be obtained from (\ref{FA1}) and (\ref{FA2}) by setting $P=1$. 
In view of the smallness of $\phi_s^{\rm NP}$ the results for the three 
observables are very close to the SM values. Still this example shows 
that even if no departures from SM expectation will be found in $\Bsmumu$ 
this does not necessarily mean that there is no NP around as with left-right 
symmetric couplings this physics cannot be seen in this decay except for 
small effects from the $B_s$ mixing phase. This physics could then be seen 
in $B\to K^*\mu^+\mu^-$ and $B\to K\mu^+\mu^-$ and $b\to s\nu\bar\nu$ transitions as demonstrated in Ref.~\cite{Buras:2012jb}.

In the ALRS scheme
NP contributions to $B_{s}\to\mu^+\mu^-$ enter again with full power.
Therefore the three observables in (\ref{trio})
offer as in the LHS and RHS schemes good test of NP.
In fact, as found in Ref.~\cite{Buras:2012jb}, after the $\Delta B=2$ constraints 
are taken into account the pattern of NP contributions is similar to LHS 
scheme except that the effects are smaller because the relevant 
couplings have to be smaller in the presence of LR operators in 
 $\Delta B=2$ in order to agree with the data on $\Delta M_s$. Therefore 
we will not show the plots corresponding to Figures~\ref{fig:Z1} and \ref{fig:Z2}.

\subsection{Tree-Level Neutral (Pseudo)Scalar Exchange}
\subsubsection{Basic Formulae}
We will next consider tree-level pseudoscalar or scalar exchanges that 
one encounters in various models either at the fundamental level or in an 
effective theory.
 We will denote by $H$ any spin 0 particle, and will refer specifically to a scalar or pseudoscalar as $H^0$ or $A^0$, respectively. 
It could in principle be the SM Higgs boson, but as the recent analysis in Ref.~\cite{Buras:2013rqa} shows, 
once the constraints from $\Delta F=2$ processes are taken into account, NP effects in $B_s\to \mu^+\mu^-$ through a tree-level SM Higgs exchange are 
at most $8\%$ of the usual SM contribution and hardly measurable.
The SM Higgs coupling to muons is simply too 
small. Therefore, what we have in mind here is
a new heavy scalar or pseudoscalar boson encountered in 2HDM or 
supersymmetric models. Yet, in this subsection we will make the working 
assumption that either a neutral scalar or pseudoscalar tree-level exchange 
dominates NP contributions. A general analysis of FCNC processes within 
such scenarios has been recently presented in Ref.~\cite{Buras:2013rqa}. Also 
the observables in (\ref{trio}) have been analysed there, but with the 
emphasis put 
on 
their correlations with $\Delta F=2$ observables, in particular 
$S_{\psi\phi}$. Here we will complement this study by computing the correlations 
among $\overline{R}$, $\mathcal{A}^{\mu\mu}_{\Delta\Gamma}$, 
and ${\cal S}_{\mu\mu}$, while taking the constraints from $\Delta F=2$ 
observables computed in Ref.~\cite{Buras:2013rqa} into account.

We define the flavour violating couplings of $H$ 
as follows
\be\label{eq:3.15}
 \mathcal{L}_\text{FCNC}(H)=\left[\Delta_L^{sb}(H)(\bar s P_L b)+
                      \Delta_R^{sb}(H)(\bar s P_R b)\right] H
\ee
where $\Delta_{L,R}^{sb}(H)$ are generally complex.  
Muon couplings $\Delta_{L,R}^{\mu\bar\mu}(H)$ are defined in a similar way.
Note that through (\ref{recall}) in LHS and RHS schemes only $\Delta_R^{sb}(H)$ and $\Delta_L^{sb}(H)$ are non-vanishing, respectively.

Then the relevant non-vanishing Wilson coefficients are given as follows
\begin{align}
 m_b(M_H)\sin^2\theta_W C_S &= \frac{1}{g_{\text{SM}}^2}\frac{1}{ M_H^2}\frac{\Delta_R^{sb}(H)\Delta_S^{\mu\bar\mu}(H)}{V_{ts}^* V_{tb}},\label{CS}\\
 m_b(M_H)\sin^2\theta_W C_S^\prime &= \frac{1}{g_{\text{SM}}^2}\frac{1}{ M_H^2}\frac{\Delta_L^{sb}(H)\Delta_S^{\mu\bar\mu}(H)}{V_{ts}^* V_{tb}},\label{CSP}\\
 m_b(M_H)\sin^2\theta_W C_P &= \frac{1}{g_{\text{SM}}^2}\frac{1}{ M_H^2}\frac{\Delta_R^{sb}(H)\Delta_P^{\mu\bar\mu}(H)}{V_{ts}^* V_{tb}},\label{CP}\\
 m_b(M_H)\sin^2\theta_W C_P^\prime &= \frac{1}{g_{\text{SM}}^2}\frac{1}{ M_H^2}\frac{\Delta_L^{sb}(H)\Delta_P^{\mu\bar\mu}(H)}{V_{ts}^* V_{tb}},\label{CPP}
\end{align}
where we have introduced
\begin{align}\begin{split}\label{equ:mumuSPLR}
 &\Delta_S^{\mu\bar\mu}(H)= \Delta_R^{\mu\bar\mu}(H)+\Delta_L^{\mu\bar\mu}(H),\\
&\Delta_P^{\mu\bar\mu}(H)= \Delta_R^{\mu\bar\mu}(H)-\Delta_L^{\mu\bar\mu}(H).\end{split}
\end{align}
Note that $m_b$ has to be evaluated at $\mu=M_H$.

From the hermiticity of the relevant Hamiltonian one can show that 
$\Delta_S^{\mu\bar\mu}(H)$ is real and $\Delta_P^{\mu\bar\mu}(H)$ purely imaginary. 
For convenience we define
\begin{equation}
   \Delta_P^{\mu\bar\mu}(H) \equiv i \tilde\Delta_P^{\mu\bar\mu}(H),
   \label{realA0coupl}
\end{equation}
so that $\tilde\Delta_P^{\mu\bar\mu}(H)$ is real.

Already at this stage it is instructive to see how different scenarios 
introduced in Section~\ref{sec:scenarios} are realized in this case.

\subsubsection*{Scenario A:}

In this scenario $S=0$. This can be realized simplest by setting 
$\Delta_S^{\mu\bar\mu}(H)=0$, which, through (\ref{equ:mumuSPLR}), implies that 
$\Delta_P^{\mu\bar\mu}(H)$ must be non-vanishing if the quark couplings differ 
from zero. The second possibility are left-right symmetric coupling to quarks 
but this would automatically imply also the vanishing of pseudoscalar couplings
giving $P=1$. 

\subsubsection*{Scenario B:}

We have just seen how this scenario can be obtained as a limiting case of 
{\it scenario A}. In order to have non-vanishing $S$ in this case this is 
realized by setting $\Delta_P^{\mu\bar\mu}(H)=0$, which through (\ref{equ:mumuSPLR}) implies that 
$\Delta_S^{\mu\bar\mu}(H)$ must be non-vanishing if the quark couplings differ 
from zero.

\subsubsection*{Scenario C:}

To achieve this scenario the scalar coefficients should be equal up to a sign to the pseudoscalar ones. 
This requires the exchanged spin-0 particle to be a mixed scalar--pseudoscalar state, which is beyond the scope of the present analysis.
We will instead realise {\it Scenario C} in Section~\ref{sec:SnP} by considering the presence of both a scalar and a pseudoscalar with equal masses and equal couplings to quarks.

\subsubsection*{Scenario D:}

Because a single scalar or pseudoscalar allows only $S$ or $P$ to deviate from its SM value, respectively, the intended usage case of this scenario, namely arbitrary but real valued $S$ and $P$, cannot be realised.

\subsubsection*{Scenario E:}

In this concrete model in which there are no NP contributions to 
$C_{10}^{(')}$ the vanishing of $P$ implies:
\be\label{SE}
\frac{m_{B_s}^2}{2m_\mu}\left(\frac{m_b}{m_b + m_s}\right) \left( \frac{C'_{P} - C_{P}}{C_{10}^{\rm SM}}\right)=1.
\ee
We will investigate whether this condition is consistent with existing constraints when the relevant Wilson coefficients are given as in (\ref{CP}) and 
(\ref{CPP}).

\subsubsection{Numerical Analysis}

\begin{figure}[t]
\begin{center}
\includegraphics[width=0.45\textwidth] {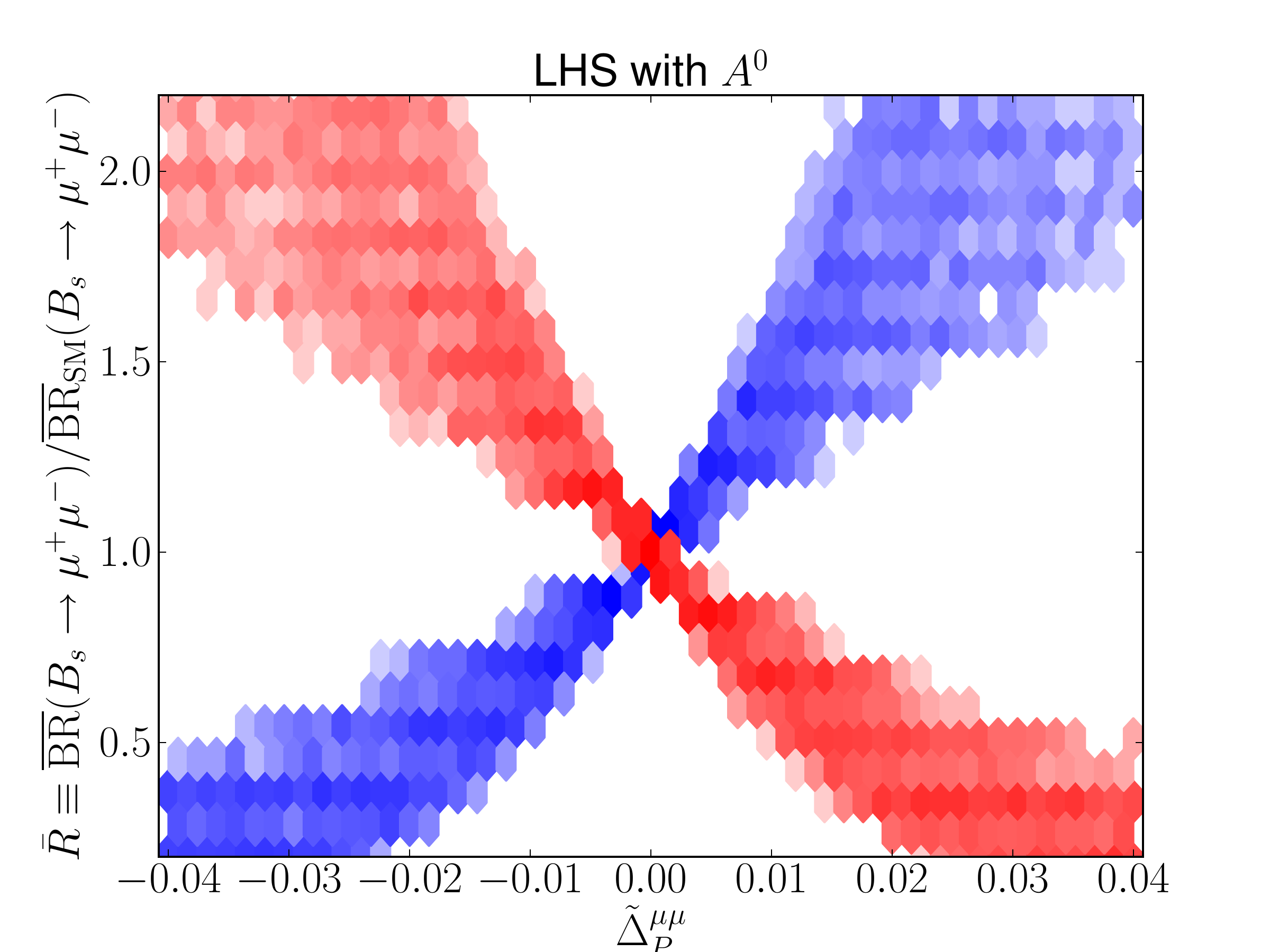}
\includegraphics[width=0.45\textwidth] {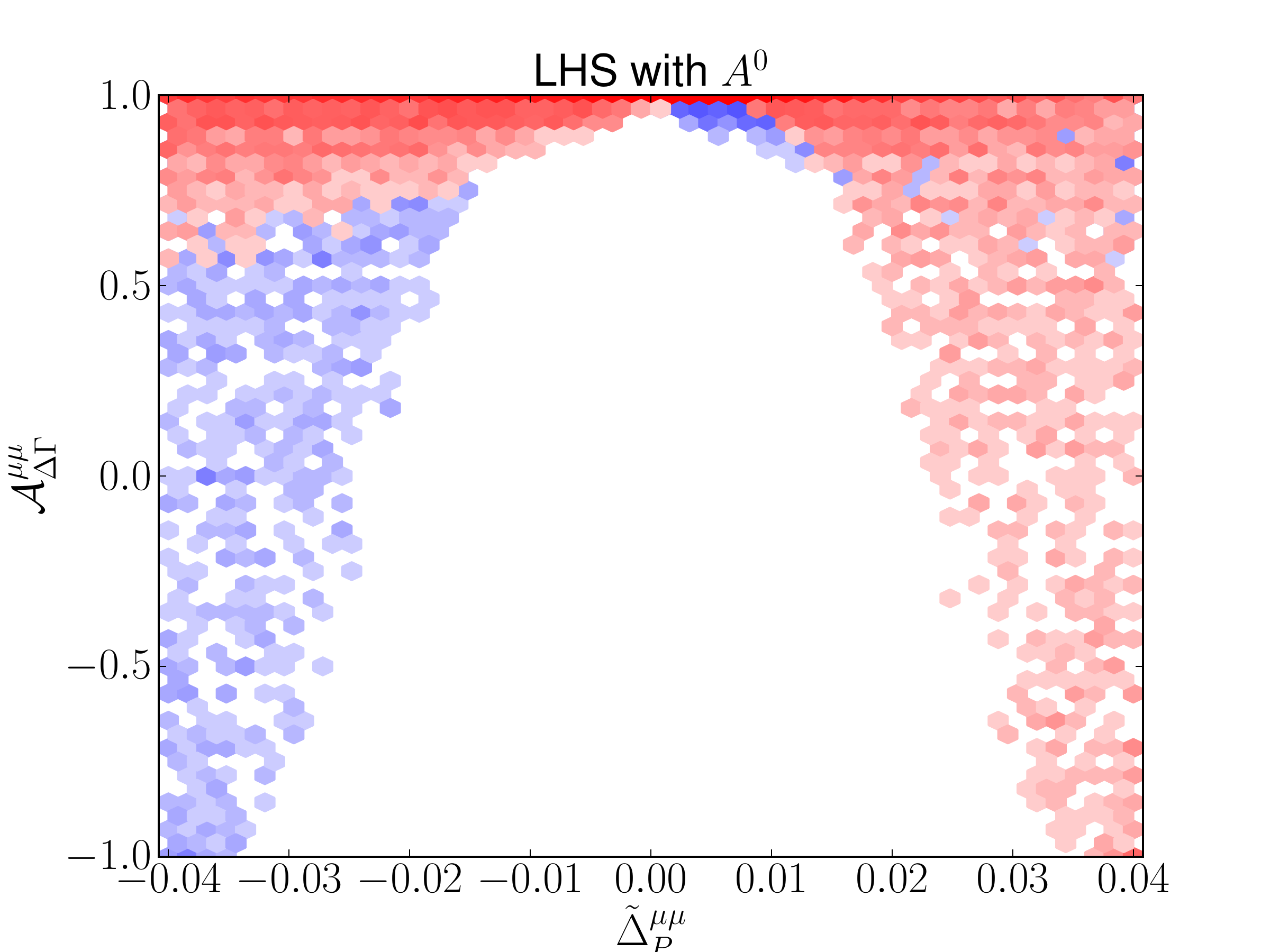}
\end{center}
\caption{The dependence of the observables $\overline{R}$ (left) and ${\cal A}_{\Delta\Gamma}^{\mu\mu}$ (right) on the pseudoscalar lepton coupling $\tilde\Delta_P^{\mu\bar \mu}(H)$ satisfying the $B_s$ mixing constraints in the LHS case. For a pseudoscalar with mass $M_{A^0} = 1\tev$.
}
\label{fig:A0leptonCouplings}
\end{figure}

Analogous to  the case of tree-level $Z'$ exchanges we will use the 
results of the $\Delta F=2$ analysis in Ref.~\cite{Buras:2013rqa} to constrain 
the quark-scalar couplings in the schemes LHS, RHS, LRS and ALRS by imposing
the conditions in (\ref{C1}).
The next step is to set values for the scalar and pseudoscalar muon couplings.
For a single scalar particle $H^0$, the parameter $|S|$ driving NP ({\it Scenario B}) is directly proportional to the muon coupling $|\Delta_S^{\mu\mu}(H^0)|$.
However, for a single pseudoscalar particle $A^0$, the muon coupling $\Delta_P^{\mu\mu}(A^0)$ is not directly proportional to $P$,
and the resulting NP observables thereby have a more involved dependence on it.
In Figure~\ref{fig:A0leptonCouplings} we show the dependence of the observables $\overline{R}$ (left panel) and ${\cal A}_{\Delta\Gamma}^{\mu\mu}$ (right panel) with respect to muon coupling $\tilde\Delta_P^{\mu\mu}(A^0)$ defined in \eqref{realA0coupl} satisfying the $B_s$ mixing constraints for the LHS case.
We observe that the parameter space of the NP physics observables is very dependent on whether we pick a large or small coupling, and that a fixed coupling cannot do it justice.
We further observe that the oases become indistinguishable if the sign of the coupling is not fixed.

In order to compare the oases behaviour of the scalar and pseudoscalar we begin by fixing the muon couplings to
\be
\label{leptonicset}
\Delta_S^{\mu\bar\mu}(H^0)=0.024, \qquad
\Delta_P^{\mu\bar\mu}(A^0)= i~ 0.012,
\ee
and $\Delta_P^{\mu\bar\mu}(H^0)=\Delta_S^{\mu\bar\mu}(A^0)=0$.

As  demonstrated in Ref.~\cite{Buras:2013rqa} these values are consistent with the allowed range for $\mathcal{B}(B_s\to\mu^+\mu^-)$ when
the constraints on the quark couplings from $B_s^0-\bar B_s^0$ are 
taken into account and $M=1\tev$. All other input parameters are as in
Ref.~\cite{Buras:2013rqa}.
The reason for choosing the scalar couplings to be larger than the pseudoscalar 
ones is that they are more weakly constrained than the latter because the 
scalar contributions do not interfere with SM contributions. The constraints from 
$b\to s\ell^+\ell^-$ transitions do not have any impact in the (pseudo) scalar 
case as shown in Ref.~\cite{Buras:2013rqa}.

\begin{figure}[t]
\begin{center}
\includegraphics[width=0.45\textwidth] {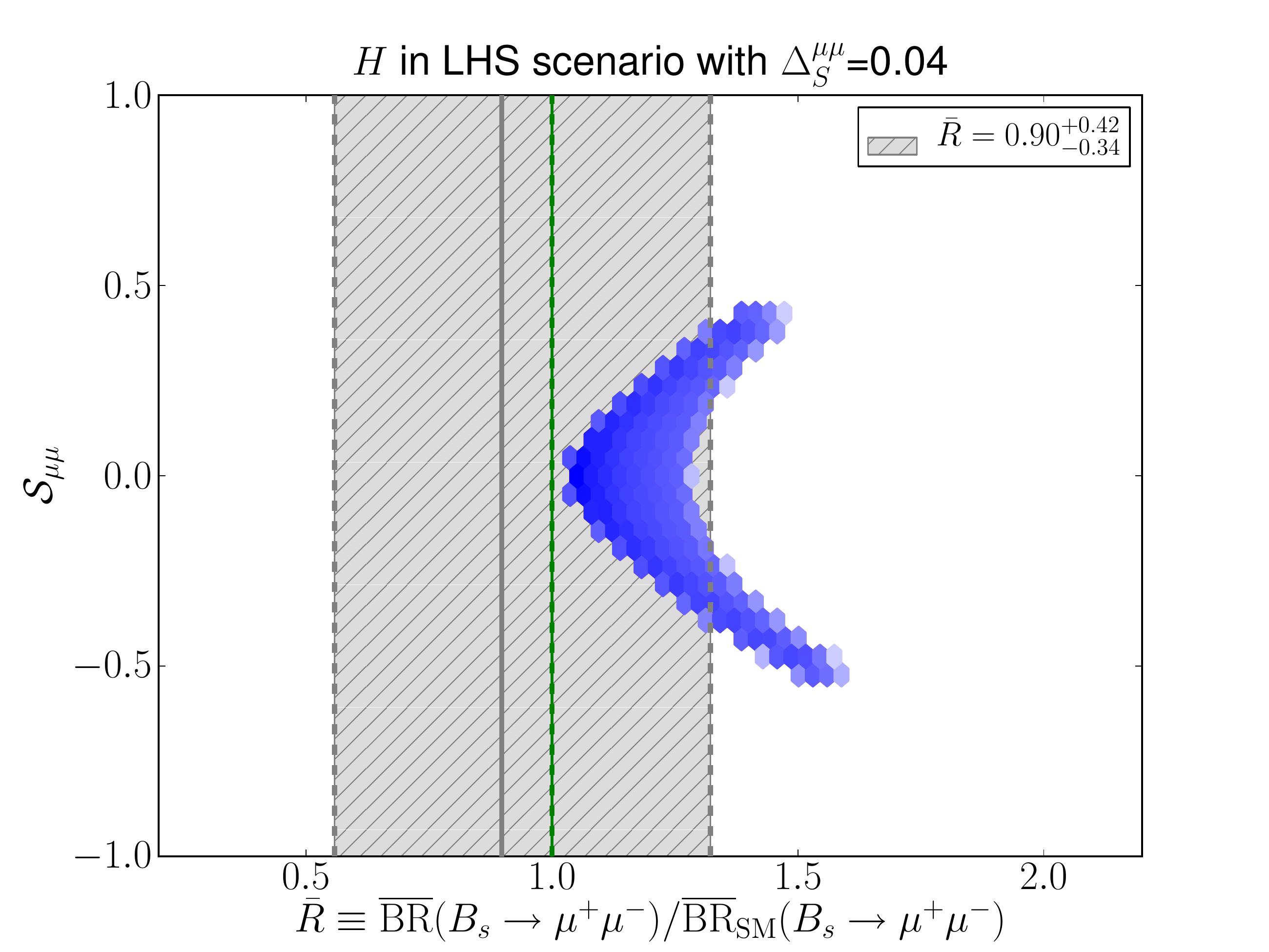}
\includegraphics[width=0.45\textwidth] {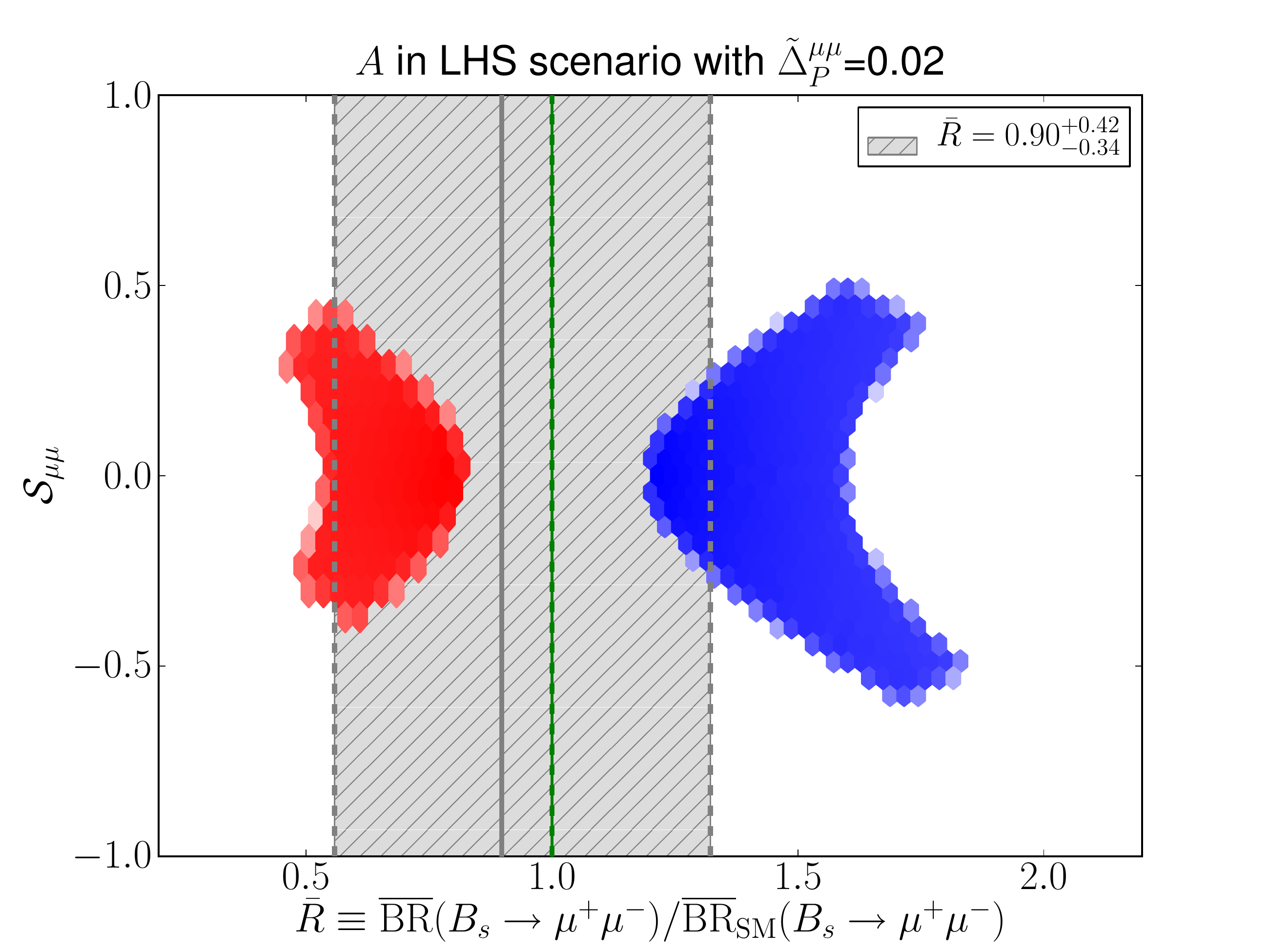}
\end{center}
\caption{
$\mathcal{S}_{\mu\mu}$  versus $\overline{R}$ 
for for LHS scheme with a scalar (left) and pseudoscalar (right) for
$M_{H^0} = M_{A^0} = 1\tev$.  Gray region: exp 1$\sigma$ range for  
$\overline{R}$.
}
\label{fig:H2}
\end{figure}

In Figure~\ref{fig:H2} we show the correlations of $\mathcal{S}_{\mu\mu}$ versus $\overline{R}$ satisfying $B_s$ mixing constraints for a single tree-level scalar (left) and pseudoscalar (right) exchange in the LHS scheme.
For the scalar case the blue and red oases overlap.
The red oases in the pseudoscalar case corresponds to $\overline{R} < 1$ and is therefore clearly distinguishable from the scalar case, where $\overline{R} > 1$ for both oases.
In Section~\ref{sec:ZprimeCompare} we will compare these correlation with $Z'$ exchange.

\begin{figure}[t]
\begin{center}
\includegraphics[width=0.45\textwidth] {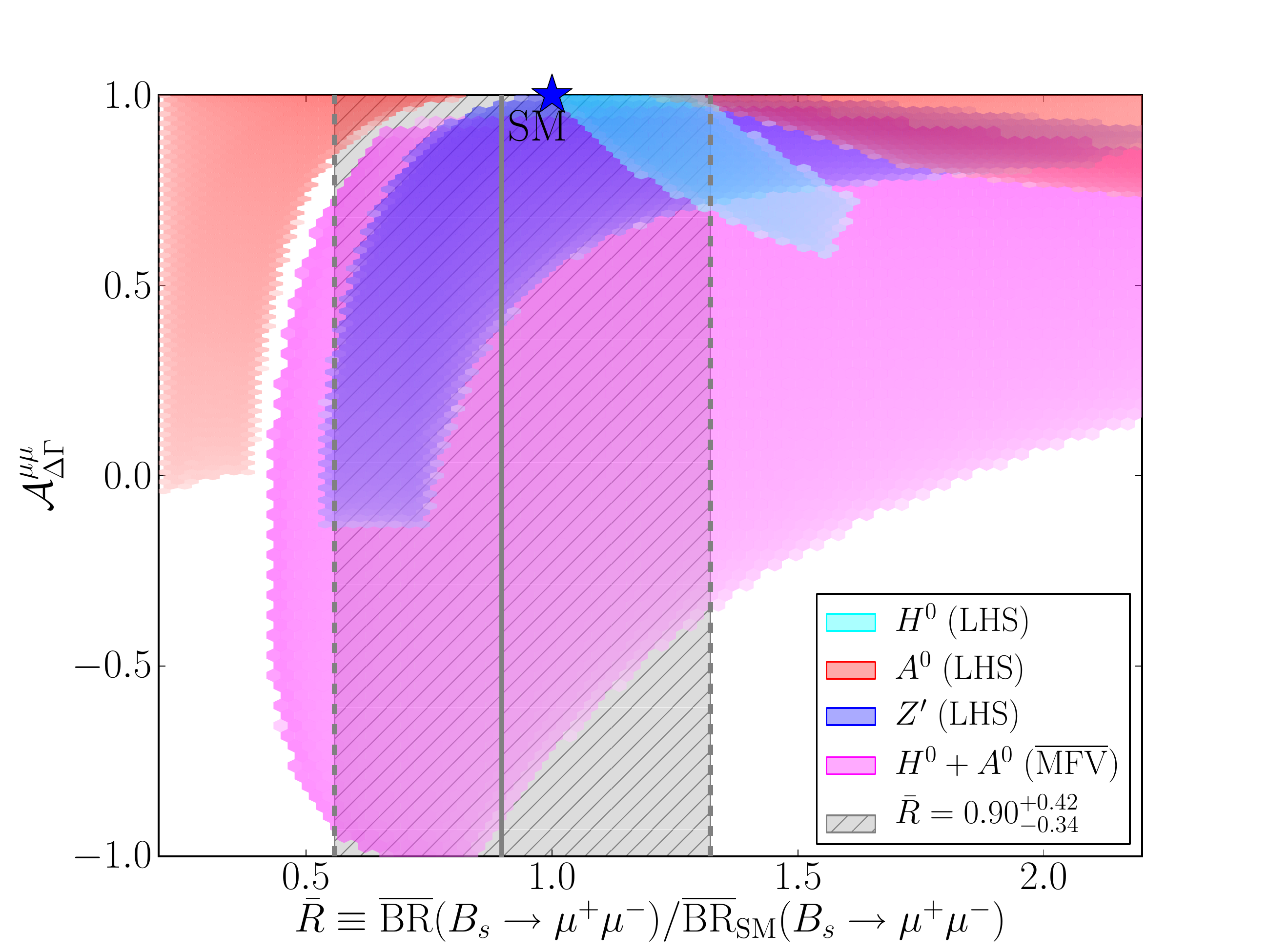}
\includegraphics[width=0.45\textwidth] {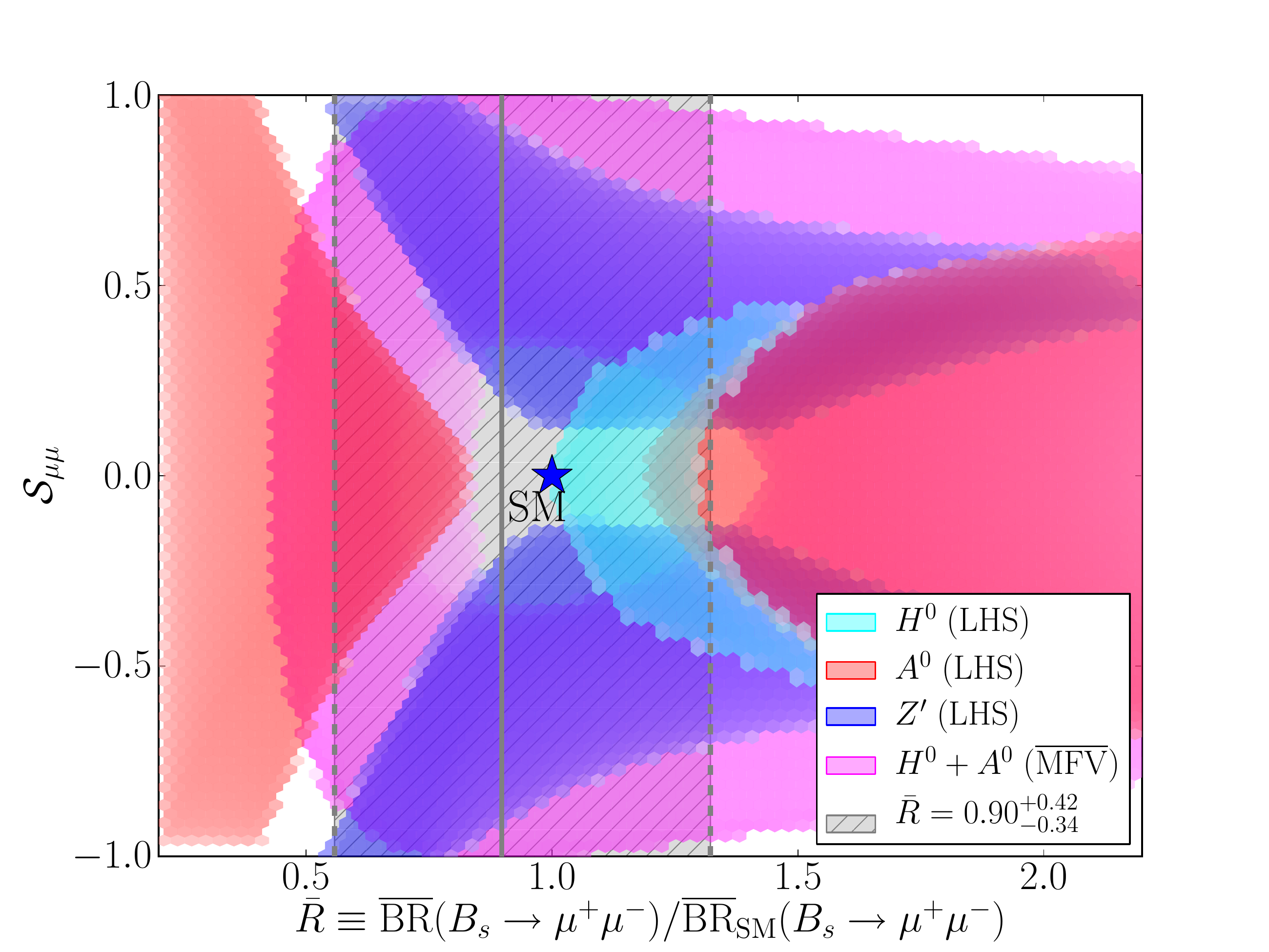}
\end{center}
\caption{
Overlay of the correlations for $\overline{R}$ versus ${\cal A}^{\mu\mu}_{\Delta\Gamma}$ (left) and ${\cal S}_{\mu\mu}$ (right) for the various specific models considered.
The lepton couplings are varied in the ranges $|\Delta_{S,P}^{\mu\mu}(H)| \in [0.012,0.024]$ and $\Delta_A^{\mu\mu}(Z')\in [0.3,0.7]$.
All particles are taken to have a mass of $1\tev$.
}
\label{fig:GRAND1}
\end{figure}

As we stated earlier, fixing the pseudoscalar muon couplings to one value does not reveal the full structure of the NP parameter space.
We therefore now consider the muon couplings varied over the following range:
\begin{equation}
    |\Delta_S^{\mu\bar\mu}(H^0)|,\ |\Delta_P^{\mu\bar\mu}(A^0)| \in [0.012,0.024].
    \label{leptonvary}
\end{equation}
From here on we will ignore the sign of the lepton couplings, but again note that this degeneracy can be resolved in the pseudoscalar case if the blue or red oasis from the $B_s$ mixing constraints can be singled out.
We thus also stop distinguishing between the two oases.

In Figure~\ref{fig:GRAND1}, $\overline{R}$ is plotted against ${\cal A}^{\mu\mu}_{\Delta\Gamma}$ (left panel) and ${\cal S}_{\mu\mu}$ (right panel) with the regions allowed by the $B_s$ mixing constraints overlayed for the various specific tree-level models discussed in this section.
The scalar and pseudoscalar muon couplings have been varied as just discussed,
and also the $Z'$ muon couplings have been varied: $\Delta_A^{\mu\mu}(Z')\in [0.3,0.7]$.
These three models are shown for the LHS scheme.
For the $Z'$ model the allowed region has also been constrained by a $2\,\sigma$ CL combined fit of the Wilson coefficient $C_{10}$ from $b\to s l^+l^-$ transitions~\cite{Altmannshofer:2012ir}.
Focusing for the moment on models with single scalar ($H^0$) or pseudoscalar ($A^0$) exchanges, the following observations can be made:
\begin{itemize}
\item
 The branching ratio observable $\overline{R}$ can in the scalar case only be enhanced
 as there is no interference with the SM contribution. On the other hand,
 in the pseudoscalar case it can be suppressed or enhanced depending on which 
 oasis of parameters is chosen. 
\item 
The values of 
 $\mathcal{A}^{\mu\mu}_{\Delta\Gamma}$ are positive for both  $H^0$ and $A^0$  
 and for $\overline{R}$ within one $\sigma$ experimental value close 
to unity.
\item
 ${\cal S}_{\mu\mu}$ can reach $\pm0.50$ in both cases. 
\end{itemize}

We do not show the corresponding results in the RHS scheme as, similarly to
the gauge boson case, the correlations in question have identical structure
with the following difference between scalar and pseudoscalar cases 
originating in the absence and presence of correlation with SM contributions, 
respectively:
\begin{itemize}
\item
    In the scalar case the two correlations in question are invariant under 
the change of LHS to RHS.
\item
    In the pseudoscalar case the structure of two correlations remain but 
going from LHS to RHS the colours have to be interchanged as was the 
case for gauge bosons.
\end{itemize}

In the LRS case, as expected, NP effects are very small as scalar and pseudoscalar contributions are absent and (\ref{FB1})
applies.
We then find for the muon couplings fixed as in \eqref{leptonicset}:  
\be 
0.984\le {\cal A}^{\mu\mu}_{\Delta\Gamma}\le 1.00, \qquad
	|{\cal S}_{\mu\mu}|\le 0.18.
\ee

Finally we investigated whether the relation (\ref{SE}), representing 
{\it Scenario E} is still consistent with all available constraints. 
This is not the case if we take the pseudoscalar lepton coupling chosen in \eqref{leptonicset} and a mass for the pseudoscalar of 1\,TeV.
For the LHS and RHS schemes a lepton coupling of $\Delta_P^{\mu\bar\mu}(H)\approx\pm i~ 0.06$ is needed to satisfy the relation.
If a pseudoscalar does manage to make $P$ vanish, then a scalar particle is needed to satisfy the lower bound on $\overline{R}$.
Such a model, with both a pseudoscalar and scalar particle present, is discussed in Section~\ref{sec:SnP}.

\boldmath
\subsubsection{Comparison with $Z'$ Scenario\label{sec:ZprimeCompare}}
\unboldmath
While the discussion presented above shows that the contributions of scalars 
and pseudoscalars can be distinguished through the observables considered, 
more spectacular differences occur when one includes the $Z'$ scenario in this 
discussion. 
Indeed the correlation between ${\cal S}_{\mu\mu}$ and $\overline{R}$ in the left panel of Figure~\ref{fig:Z1} has a very different structure from the case of pseudoscalar or scalar exchanges shown in Figure~\ref{fig:H2}. 

    In the right panel of Figure~\ref{fig:GRAND1} an overlay of these regions is shown for LHS schemes, with the lepton couplings varied as given in \eqref{leptonvary}.
Similarly, in the left panel of Figure~\ref{fig:GRAND1} we show the correlation between ${\cal A}^{\mu\mu}_{\Delta\Gamma}$ and $\overline{R}$, where strong contrasts between the allowed regions also emerge.
The difference between the $Z'$ and pseudoscalar exchange is striking because, unlike for a scalar,
both particles generate {\it Scenario A}.

The difference between the $A^0$-scenario and $Z'$-scenario in 
question can be traced back to the difference between the phase of the NP correction to $\tilde{P}$, which was defined in \eqref{PTilde}. 
As the  phase $\delta_{23}$  in the quark coupling $\Delta_L^{bs}$ from the analysis of $B_s$-mixing in Ref.~\cite{Buras:2013rqa} is  the
same in both scenarios the difference enters 
through the muon couplings, which are imaginary in the case of the $A^0$-scenario but 
real in the case of $Z'$.  
This is why the structure 
of correlations in both scenarios is so different. Taking also 
the sign difference between $Z'$ and pseudoscalar contributions to 
the $b\to s \mu^+\mu^-$ amplitude into account, we find that 
\be\label{PP}
P(Z')=1+ r_{Z'} e^{i \delta_{Z'}}, \qquad P(A^0)=1 + r_{A^0} e^{i\delta_{A^0}}
\ee
with 
\be\label{SHIFT}
r_{Z'}\approx r_{A^0}, \qquad    \delta_{Z'}=\delta_{23}-\beta_s, \qquad \delta_{A^0}=\delta_{Z'}-\frac{\pi}{2}.
\ee
It turns out that the $B_s$ mixing constraints force the phase $ \delta_{23}$
to be in the ballpark of
$90^\circ$ and $270^\circ$ for the blue and red oasis, respectively \cite{Buras:2012jb,Buras:2013rqa}. 
This implies in the case of the $Z'$ scenario, as seen in Figure~\ref{fig:Z1}, 
positive and negative value of ${\cal S}_{\mu\mu}$  for the blue and 
red oasis, respectively. Simultaneously  $\overline{R}$, where NP is governed by $\cos\delta_{Z'}$, can be enhanced or suppressed 
in each oasis. On the other hand, (\ref{SHIFT}) implies that the
phase $\delta_{A^0}$ is in the ballpark of $0^\circ$ and $180^0$ for the 
blue and red oasis, respectively.  Therefore the asymmetry 
${\cal S}_{\mu\mu}$ can vanish in both oases, while this was not possible in 
the $Z'$ case. As NP in $\overline{R}$ is governed by 
$\cos\delta_{A^0}$, this enhances and suppresses $\overline{R}$ for blue and red 
oasis, respectively as clearly seen in Figure~\ref{fig:H2}. In particular,
$\overline{R}$ differs from its SM value, while this is not the case in the $Z'$ 
scenario. Finally, let us note that with larger values of muon 
couplings NP effects in $\overline{R}$, $\mathcal{A}^{\mu\mu}_{\Delta\Gamma}$ and 
${\cal S}_{\mu\mu}$ can be larger than shown in Figs.~\ref{fig:Z1} and \ref{fig:Z2} (see, for example, Figure~\ref{fig:GRAND2}).

What is particularly interesting is that
 these differences 
are directly related to the difference in the fundamental properties of the particles involved: their spin and CP-parity. As far as the last property is 
concerned, also differences between the implications of the pseudoscalar and scalar exchanges have been identified as discussed in detail above. They are 
related to the fact
that the scalar contribution, being $CP$ even, cannot interfere with the SM contribution.

\boldmath
\subsection{Tree-Level Neutral Scalar+Pseudoscalar Exchange\label{sec:SnP}}
\unboldmath
\subsubsection{Basic Formulae}

In this model we assume the presence of a scalar $H^0$ and  pseudoscalar $A^0$ with equal (or nearly degenerate) mass $M_H$.
This is, for example, effectively realised in 2HDMs in a decoupling regime, where $H^0$ and $A^0$ are much heavier than the SM Higgs $h^0$ and almost degenerate in mass~\cite{Gunion:2002zf}.
{We will show that under specific assumptions this setup can reproduce {\it Scenarios C},  {\it D} or {\it E}.}

The couplings of the scalar and pseudoscalar to quarks are given in general by the following flavour-violating Lagrangian:
\begin{align}
    {\cal L}_{\rm FCNC}(H^0,A^0)
    =& \left[\Delta_L^{sb}(H^0)(\bar s P_L b)+
                      \Delta_R^{sb}(H^0)(\bar s P_R b)\right] H^0 \notag\\
    &+ \left[\Delta_L^{sb}(A^0)(\bar s P_L b)+
                      \Delta_R^{sb}(A^0)(\bar s P_R b)\right] A^0.
                      \label{HnALagFCNC}
\end{align}
We will assume that the scalar and pseudoscalar couple with equal strength to quarks:
\begin{equation}
    {\cal L} \ni \bar D_L \tilde\Delta D_R (H^0 + iA^0) + {\rm h.c}, 
    \label{HnALagEqual}
\end{equation}
where $D=(d,s,b)$ and $\tilde\Delta$ is a matrix in flavour space.
Then
\begin{align}
    \Delta^{sb}_R(H^0) &= \tilde\Delta^{sb}, & \Delta^{sb}_L(H^0) &= \left[\tilde\Delta^{bs}\right]^*, \notag\\  
    \Delta^{sb}_R(A^0) &= i\tilde\Delta^{sb}, & \Delta^{sb}_L(A^0) &= -i\left[\tilde\Delta^{bs}\right]^*.
\end{align}
where in general $\tilde\Delta^{sb},\tilde\Delta^{bs} \in {\mathbb C}$.

\subsubsection*{Scenario C:}

{To reproduce this scenario we set the pseudoscalar and scalar masses to be exactly equal: $M_{H^0} = M_{A^0} = M_H$.}
Further relating the lepton couplings by a single real parameter $\tilde\Delta^{\mu\bar\mu}$:
\be 
\Delta^{\mu\bar\mu}(H^0)=\tilde\Delta^{\mu\bar\mu},\qquad \Delta^{\mu\bar\mu}(A^0)=i\,\tilde\Delta^{\mu\bar\mu}
\label{leptonCouplingsEqual}
\ee
and inserting the lepton and quark couplings into formulae (\ref{CS})--(\ref{CPP}) 
we find:
\begin{align}
    C_S = -  C_P &= \frac{1}{g_{SM}^2\,M_H^2\, m_b \sin^2\theta_W}   \frac{\tilde\Delta^{sb} \tilde\Delta^{\mu\mu}}{V_{ts}^* V_{tb}}\\
    C'_S = C'_P &= \frac{1}{g_{SM}^2\,M_H^2\, m_b \sin^2\theta_W} \frac{\left[\tilde\Delta^{bs}\right]^* \tilde\Delta^{\mu\mu}}{V_{ts}^* V_{tb}}.
\label{CprimeSP}
\end{align}

This simple model satisfies the relations in (\ref{2HDM}) and thereby belongs 
to {\it Scenario C}.
These relations are in fact valid for all the quark coupling schemes: LHS, RHS, LRS and ALRS.
Yet the physics implications depend on the scheme considered: 
\begin{itemize}
\item
In LHS and RHS schemes NP contributions to $B_s^0$--$\bar B_s^0$ mixing from 
scalar and pseudoscalar with the same mass cancel each other so that there 
is no constraint from $B_s^0$--$\bar B_s^0$ mixing.
Thus NP effects in $B_s\to\mu^+\mu^-$ can only be constrained by the decay itself or other 
$b\to s\ell^+\ell^-$ transitions.
\item
In LRS and ALRS  schemes non-vanishing contributions from LR operators to 
$B_s^0$--$\bar B_s^0$ mixing are present. Moreover we find
\be
C_S=C_S^\prime, \qquad C_P=-C_P^\prime   \qquad ({\rm LRS}),
\ee
\be
C_S=-C_S^\prime, \qquad C_P=C_P^\prime   \qquad ({\rm ALRS}). 
\ee
Therefore in the LRS case only pseudoscalar contributes to $\Bsmumu$ ({\it Scenario A}), while in the ALRS case only scalar contributes ({\it Scenario B}).
\end{itemize}

We conclude therefore that in order to have an example of {\it Scenario C} that 
differs from {\it Scenario A} and {\it B} and moreover in which NP contributions to 
$B_s^0$--$\bar B_s^0$ mixing are present, we need both $L$ and $\overline{R}$ couplings which 
are not equal to each other or do not differ only by a sign.

An option to reproduce {\it Scenario C} with non-trivial constraints from mixing is given by Minimal Flavour Violation (MFV). 
In the MFV formalism $\tilde\Delta$ is constructed out of the spurion matrices $Y_U$ and $Y_D$~\cite{D'Ambrosio:2002ex}.
In principle the following constructions can contribute to the $b\to s$ FCNCs at leading order in the off-diagonal structure:
\begin{equation}
    Y_U Y_U^\dagger\, Y_D,\quad Y_D Y_D^\dagger\, Y_U Y_U^\dagger\, Y_D,\quad Y_U Y_U^\dagger\, Y_D Y_D^\dagger\, Y_D.
\end{equation}
However, the last two will in general receive dynamical (loop) suppressions.
Thus, for simplicity, we assume the first construction to be dominant.
In the notation of Ref.~\cite{Buras:2010mh}, where MFV is discussed in the context of a general 2HDM with flavour blind phases (${\rm 2HDM_{\overline{\rm MFV}}}$), this is equivalent to assuming $|a_0|\gg |a_1|,|a_2|$.
As a result we find
\begin{equation}
    \tilde\Delta^{sb}=\epsilon\, y_b\, y_t^2\, V_{ts}^*V_{tb},\qquad
    \left[\tilde\Delta^{bs}\right]^*=\epsilon^*\, y_s\, y_t^2\, V_{ts}^*V_{tb}
    = \frac{m_s}{m_b}\frac{\epsilon^*}{\epsilon}\tilde\Delta^{sb}.
    \label{MFVDeltas}
\end{equation}
Thus under the above assumptions all of the quark couplings in \eqref{HnALagFCNC} can be expressed in terms of a single NP parameter $\epsilon$ 
\footnote{It should be emphasized that in general this is not 
the case for ${\rm 2HDM_{\overline{\rm MFV}}}$ \cite{Buras:2010mh,Buras:2010zm}. See additional comments below.}.

The parameter $\epsilon$ is real in pure MFV but may be complex in 
${\rm 2HDM_{\overline{\rm MFV}}}$~\cite{Buras:2010mh}.
Inserting relation \eqref{MFVDeltas} into \eqref{CprimeSP} we find
\begin{align}
 C_S^\prime &=\frac{m_s}{m_b}\frac{\epsilon^*}{\epsilon} C_S &
C_P^\prime &=-\frac{m_s}{m_b}\frac{\epsilon^*}{\epsilon} C_P,
\end{align}
and observe a $m_s/m_b$ suppression of the primed operators.
In pure MFV, where $\epsilon$ is real, the parameters $C^{(\prime)}_{S,P}$ are also all real.

\subsubsection*{Scenario D:}

In {\it Scenario D} the parameters $P$ and $S$ are arbitrary but do not carry new CP violating phases.
The pure MFV model with a scalar and pseudoscalar that we just discussed is therefore a natural candidate.
However, because this model was defined to satisfy {\it Scenario C}, as it stands we have $P\pm S=1$.
If we continue to insist that the scalar and pseudoscalar should couple with equal strengths and phases to quarks as in \eqref{HnALagEqual}, then there are two choices for making $P$ and $S$ arbitrary.

One choice is to allow different couplings to leptons for the scalar and pseudoscalar i.e.\ $|\Delta^{\mu\bar\mu}(H^0)|\neq |\Delta^{\mu\bar\mu}(A^0)|$.
In this case the constraints from $B_s$ mixing (discussed below) do not change, and only the current bounds on $\overline{R}$ must be satisfied.

Alternatively, a non-trivial difference between the scalar mass $M_{H^0}$ and the pseudoscalar mass $M_{A^0}$ can be introduced.
In this case the lepton couplings can remain equal as defined in \eqref{leptonCouplingsEqual}.
The catch, however, is that now the LL and (to a much lesser extent in MFV) RR contributions to $B_s$ mixing no longer vanish.
Thus the allowed mass difference, and thereby the arbitrariness of $P$ and $S$ is constrained by mixing.

\subsubsection*{Scenario E:}

This scenario requires that $P=0$ and therefore that $S$ alone generates a value of $\overline{R}$ large enough to meet the current experimental bounds.
As we are dealing with two spin-0 particles, the pseudoscalar in the present model must satisfy the relation given in \eqref{SE}.

By definition {\it Scenario E} allows $S$ to have a new CP violating phase $\varphi_S$.
However, the relation in \eqref{SE} requires that ${\rm arg}(C_P - C'_P)=0$.
Therefore, because we required that the scalar and pseudoscalar should couple to quarks with equal strengths and phases (see \eqref{HnALagEqual}), it follows that $\varphi_S=0$.
The {\it Scenario E} realisable in this model is therefore just a specific case of {\it Scenario D}, where $P$ and $S$ are real and arbitrary, with the addition that $P$ is tuned to vanish.

In the next section we will address whether, given that $S$ and $P$ are made to vary due to a scalar--pseudoscalar mass difference and $P$ is tuned to zero, the range of $S$ allowed by mixing can satisfy the experimental bounds on $\overline{R}$.

\subsubsection{Numerical Analysis}

Our numerical analysis for this model will focus on the above mentioned assumptions that produce {\it Scenario C}.
Specifically, we begin by assuming an exactly degenerate scalar mass $M_H$, equal scalar and pseudoscalar lepton couplings and MFV.
At the end of this section we also briefly address the consequences of a scalar--pseudoscalar mass difference, which could produce {\it Scenarios D} and {\it E}.

By imposing MFV on the flavour matrix $\tilde\Delta$ introduced in \eqref{HnALagEqual}, it follows that the analogues of $\tilde\Delta^{sb}$ in the $B_d$ and $K$ systems are related to it by
\begin{equation}
    \tilde\Delta^{db} = - \frac{V_{td}^*}{V_{ts}^*} \tilde\Delta^{sb},\quad\quad
    \tilde\Delta^{ds} = - \frac{m_s}{m_b}\frac{V_{td}^*}{V_{tb}^*} \left[\tilde\Delta^{sb}\right]^*.
    \label{DeltaOthers}
\end{equation}
Therefore the value taken by $\tilde\Delta^{sb}$ should in principle not only satisfy the experimental $B_s$ mixing constraints, but also those of the $B_d$ and $K$ systems.
In practice, however, NP contributions in this model to $B_d$ mixing are suppressed by a factor of $m_d/m_s$ relative to $B_s$ mixing and thereby 
very small. As a result, this model with MFV cannot relieve the current tensions in $B_d$ mixing between theory and experiment \cite{Lunghi:2008aa,Buras:2008nn}.
Contributions to neutral Kaon mixing are totally negligible.
We therefore proceed to only consider constraints from $B_s$ mixing.

The only contribution that survives in $B_s$ mixing is the LR one and this introduces  the following shift  in the SM box
function~\cite{Buras:2013rqa}
\begin{equation}
    S(B_s) = S_0(x_t) + \left[\Delta S({B_s})\right]_{LR}, 
    \label{SMboxMod}
\end{equation}
where 
\begin{align}\label{LR}
    \left[\Delta S({B_s})\right]_{LR} &= 2\,r^{LR}\frac{[\tilde{\Delta}^{sb}]^*\tilde{\Delta}^{bs}}{M^2_H \left(V_{ts}V_{tb}^*\right)^2}
    = 2\,r^{LR}\,\left(\frac{m_s}{m_b}\right)\frac{|\epsilon|^2\,y_b^2\,y_t^4}{M^2_H}.
\end{align}
with $r^{LR}= -3\times 10^2\tev^2$ \cite{Buras:2013rqa}. 

The following observations should be made:
\begin{itemize}
\item
In spite of possible new flavour blind phases in the  $\overline{\rm MFV}$ scenario, these 
phases do not show up in $B_s$-mixing, so that the CP asymmetry $S_{\psi\phi}$ 
remains at its SM value, still consistent with experiment.
In Ref.~\cite{Buras:2010mh} the $\Delta B=2$ operator for a 2HDM with MFV is also found to leave flavour-blind phases unconstrained in the limit $|a_0|\gg |a_1|,|a_2|$.  
In general this is not the case for ${\rm 2HDM_{\overline{\rm MFV}}}$ and, 
as analysed in Refs.~\cite{Buras:2010mh,Buras:2010zm}, $S_{\psi\phi}$  can receive 
NP contributions. 
Note that these phases  also appear in the flavour-conserving Yukawa couplings, which contribute to  the electric dipole moments of various
atoms and hadrons by the exchange of Higgs fields. But as shown for  ${\rm 2HDM_{\overline{\rm MFV}}}$ in Ref.~\cite{Buras:2010zm} the
present upper bounds on EDMs do not yet have any impact on the 
observables considered here~\footnote{We thank Minoru Nagai for enlighting comments on these issues.}.
\item
On the contrary, $\Delta M_s$ receives a small (suppressed by $m_s/m_b$) {\it negative} contribution which is 
good as the SM value is roughly $10\%$ above its experimental value \cite{Buras:2012ts}.
This suppression due to LR operators within a MFV framework was first pointed out for the MSSM with MFV in Ref.~\cite{Buras:2002vd}.
\item
The fact that the flavour-blind phases are unconstrained through $B_s$-mixing 
allows us to obtain significant effects from them in $B_s\to\mu^+\mu^-$ observables
as we will see soon. 
\end{itemize}

For both MFV and $\overline{\rm MFV}$ we find the range:
\begin{equation}
    |\tilde{\Delta}^{sb}| \in [0.00196, 0.00530],
    \label{DeltaSbRange}
\end{equation}
for $M_H=1\,{\rm TeV}$, which, as seen in (\ref{MFVDeltas}), is  consistent with the tacit assumption that $\epsilon$ should be small.

\begin{figure}[t]
\begin{center}
\includegraphics[width=0.45\textwidth] {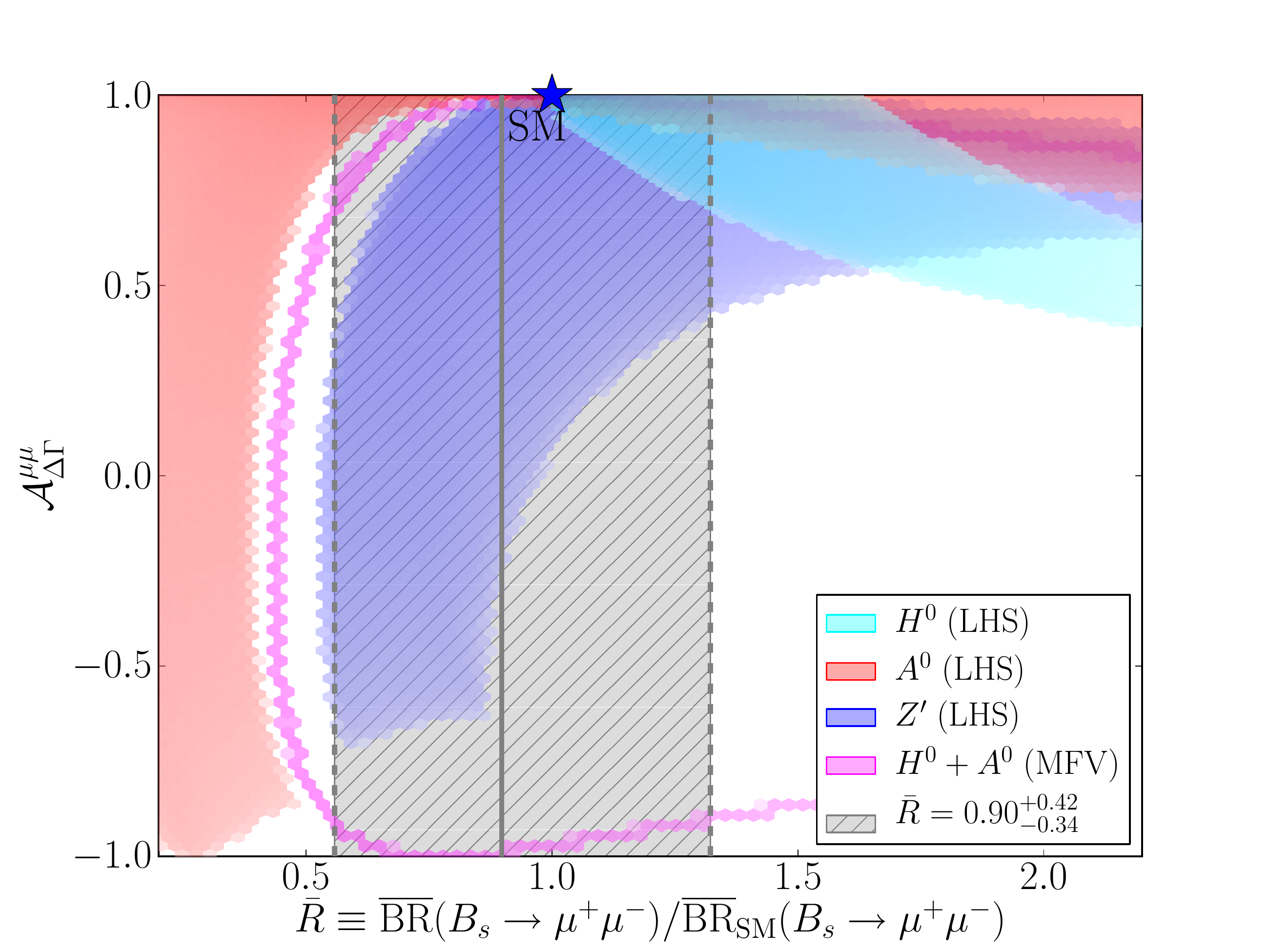}
\includegraphics[width=0.45\textwidth] {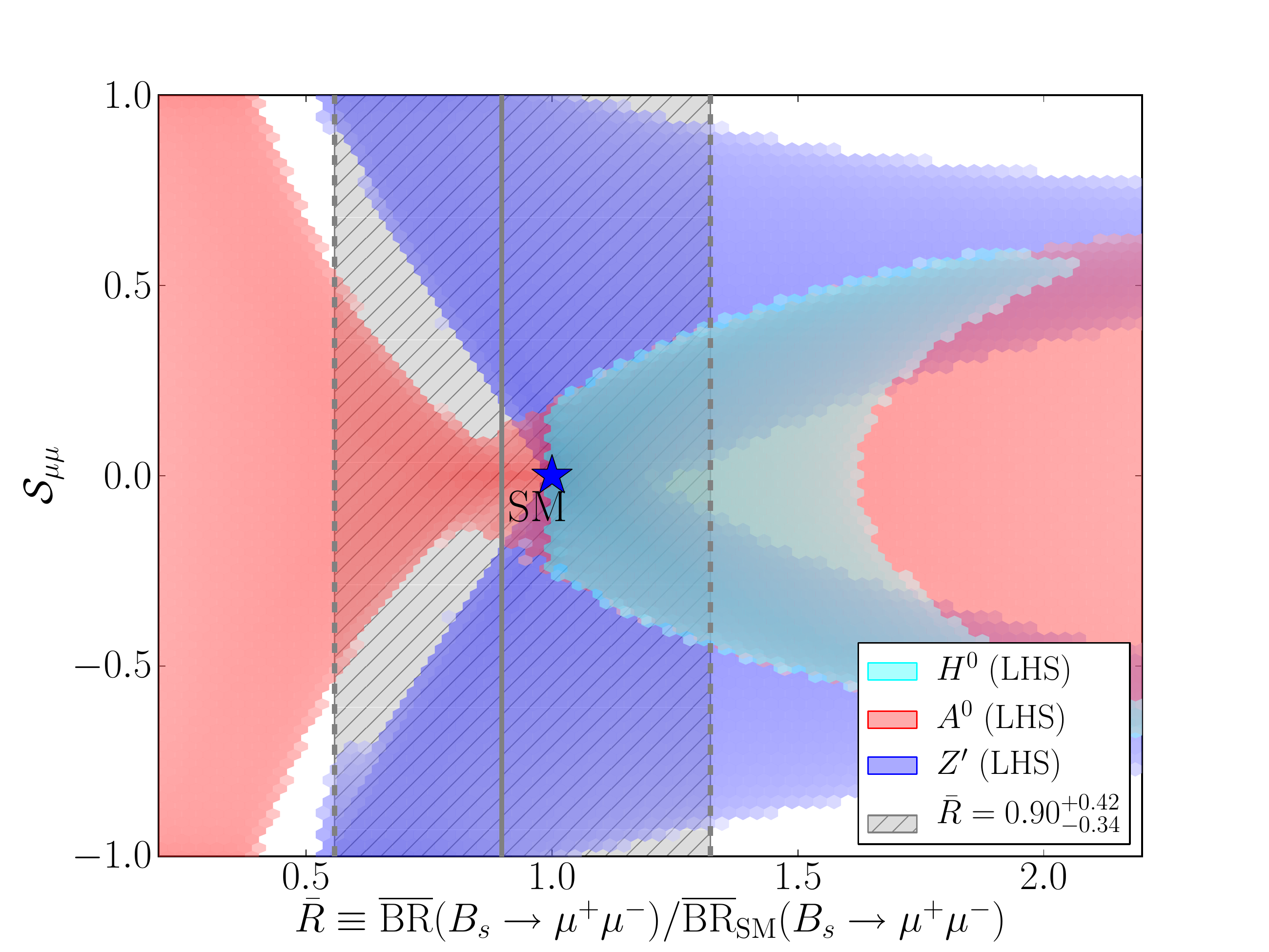}
\end{center}
\caption{
 Overlay of the correlations for $\overline{R}$ versus ${\cal A}^{\mu\mu}_{\Delta\Gamma}$ (left) and ${\cal S}_{\mu\mu}$ (right) for the various specific models considered.
The lepton couplings are varied in the ranges $|\Delta_{S,P}^{\mu\mu}(H)| \in [0.00,0.035]$ and $\Delta_A^{\mu\mu}(Z')\in [0.0,1.0]$.
All particles are taken to have a mass of $1\tev$.
}
\label{fig:GRAND2}
\end{figure}

To proceed with numerics for the $\Bsmumu$ observables we must set the coupling of $H^0$ and $A^0$ to muons.
In the context of {\it Scenario C} this means setting $\tilde\Delta^{\mu\mu}$ as defined in \eqref{leptonCouplingsEqual}.
In order to compare with the single tree-level scalar and pseudoscalar models discussed in the previous section we begin by varying the coupling between $[0.012,0.024]$ as also done in \eqref{leptonvary}.

The left panel of Figure~\ref{fig:GRAND1} shows $\mathcal{A}^{\mu\mu}_{\Delta\Gamma}$  plotted versus $\overline{R}$ for $\overline{\rm MFV}$ with $M_H =1\,{\rm TeV}$. 
The allowed region from $B_s$ mixing constraints shown in this plot should be compared with the theoretical situation sketched for {\it Scenario C} in the left panel of Figure~\ref{fig:SnP}.
By inspection of the theoretical plot one observes that the pure MFV model (with no flavour blind phases) corresponds to the outer border of the $\overline{\rm MFV}$ region shown.
It is interesting to observe that in both models a negative $\mathcal{A}^{\mu\mu}_{\Delta\Gamma}$ is possible within the $B_s$ constraints mixing, in contrast to the tree-level models considered above with a single (pseudo)scalar or gauge boson.
Because the flavour-blind phase in $\overline{\rm MFV}$ is completely unconstrained, 
almost the entire experimentally allowed region is left unconstrained by $B_s$ mixing in this model.

In the right panel of Figure~\ref{fig:GRAND1} we similarly show $\mathcal{S}_{\mu\mu}$  versus $\overline{R}$ in the $\overline{\rm MFV}$ model for $M_H =1\,{\rm TeV}$.
In the pure MFV model $\mathcal{S}_{\mu\mu}=0$ and therefore these plots are not interesting.

\begin{figure}[t]
\begin{center}
\includegraphics[width=0.45\textwidth] {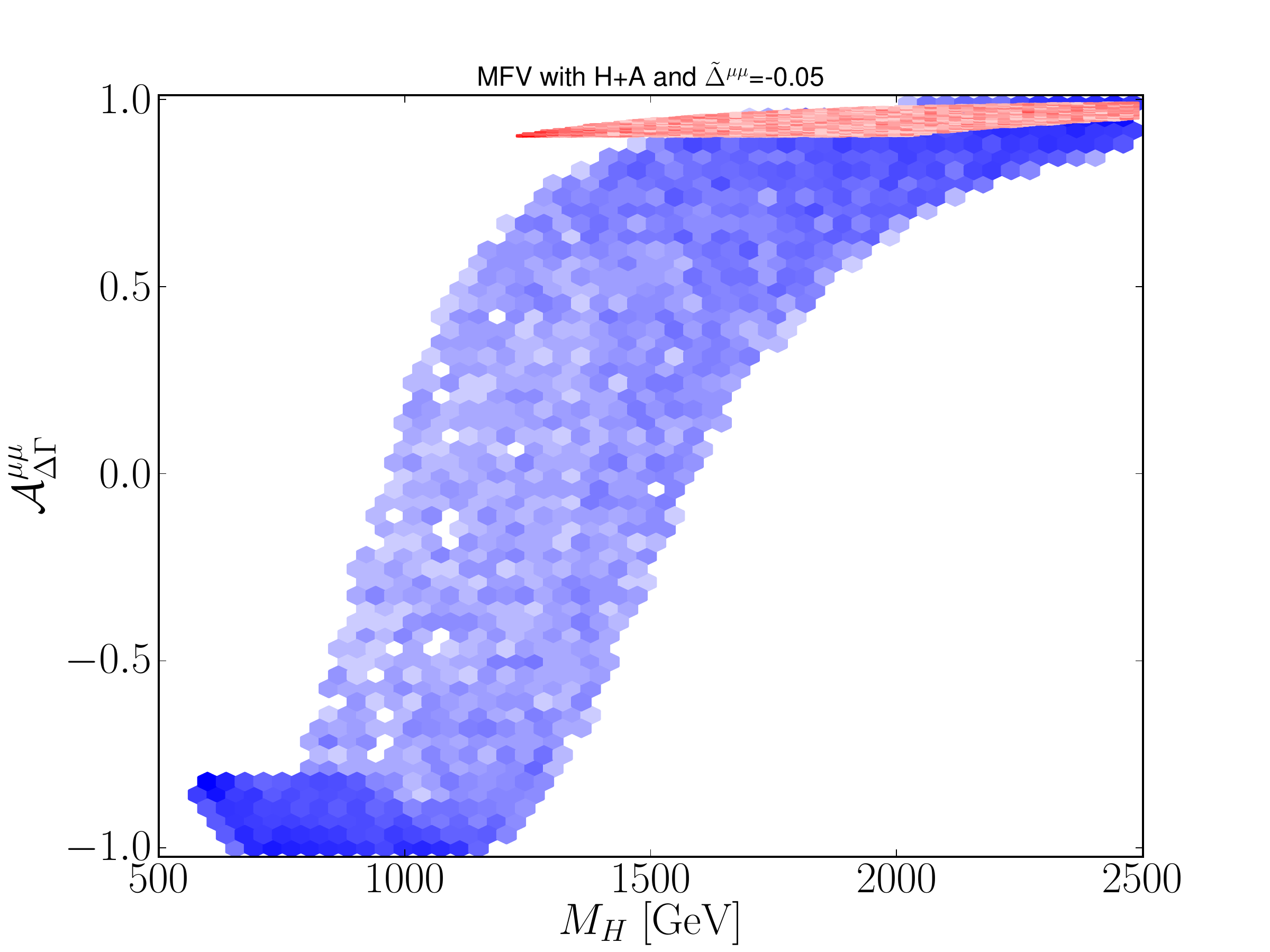}
\includegraphics[width=0.45\textwidth] {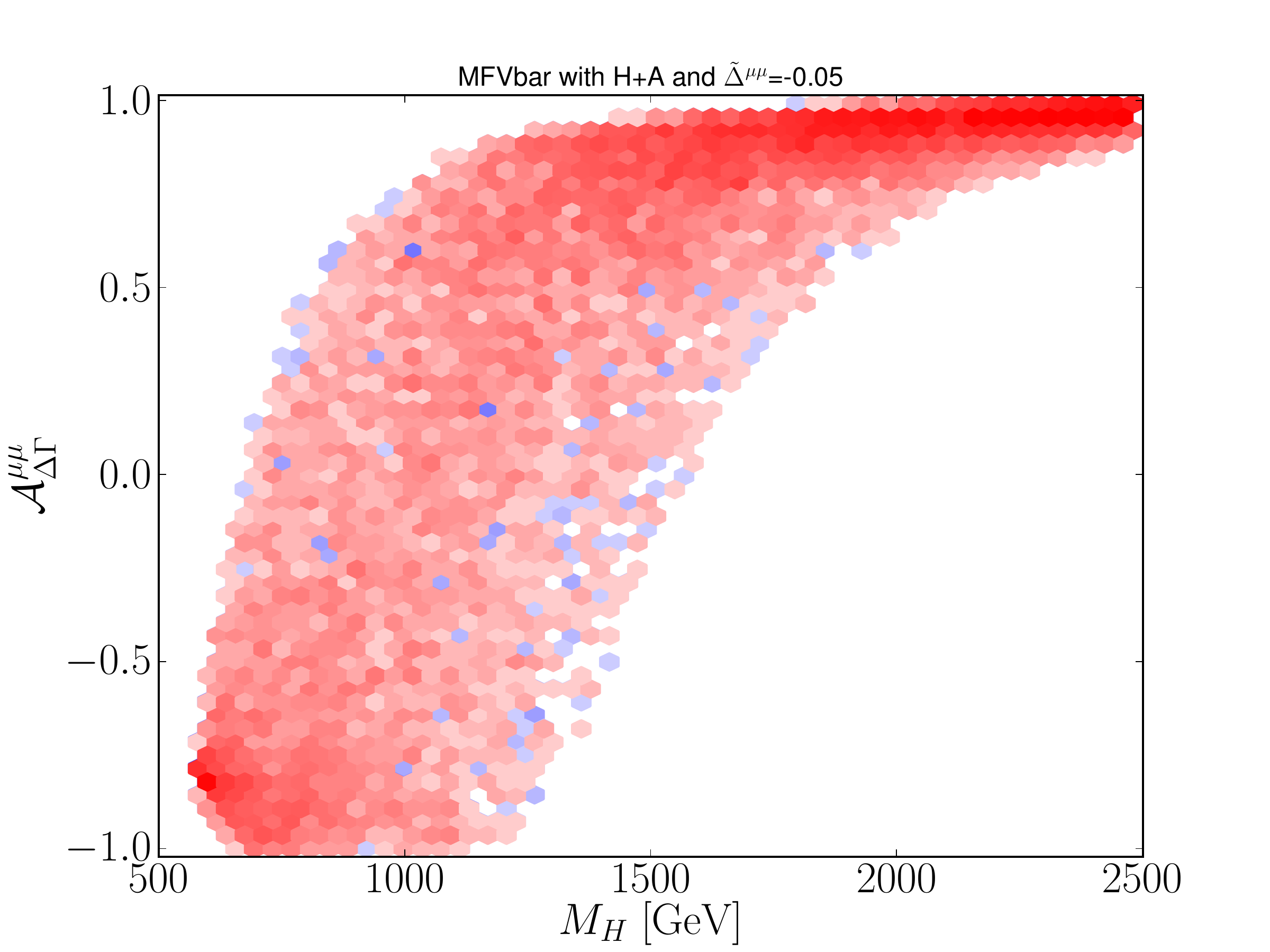}
\end{center}
\caption{
The allowed region of $\mathcal{A}^{\mu\mu}_{\Delta\Gamma}$ versus the heavy scalar mass $M_H$ in MFV (left panel) and $\overline{\rm MFV}$ (right panel).
The allowed region satisfies the $B_s$ mixing constraints and falls with the 2\,$\sigma$ C.L region of $\overline{R}$: $\overline{R} \in [0.30,1.80]$.
}
\label{fig:MH-ADG_MFV}
\end{figure}

In a 2HDM with large $\tan\beta$, which can generate a decoupled heavy scalar and pseudoscalar as discussed here, the muon coupling is given by
\begin{equation}
    \tilde\Delta^{\mu\mu} = -2\left(\frac{\sqrt{2}\,m_\mu}{v} \tan\beta\right) = -0.03 \left[\frac{\tan\beta}{25}\right],
    \label{HnALeptonCouplings}
\end{equation}
which demonstrates that the (pseudo)scalar muon couplings can be larger than what we have assumed so far.
In Figure~\ref{fig:GRAND2} we repeat the plots we have shown in Figure~\ref{fig:GRAND1}, but now with the muon couplings varied over much larger ranges:
\begin{equation}
    |\Delta_{S,P}^{\mu\bar\mu}(H^0,A^0)| \in [0.00,0.035],\qquad |\Delta_A^{\mu\bar\mu}(Z')| \in [0.0,1.0].
\end{equation}
 This range of couplings gives a better impression of the full allowed parameter space,
at the cost of hiding some of the characteristic differences between the considered models.
We do not again show the large allowed region of $H^0+A^0$ model with $\overline{\rm MFV}$, but we do show it with pure MFV in the $\mathcal{A}^{\mu\mu}_{\Delta\Gamma}$ versus $\overline{R}$ case (left panel).

In Figure~\ref{fig:MH-ADG_MFV} we show the allowed range of $\mathcal{A}^{\mu\mu}_{\Delta\Gamma}$ with respect to the heavy scalar mass $M_H$ in the MFV (left panel) and $\overline{\rm MFV}$ (right panel). 
In these plots we have fixed the muon couplings to $\tilde\Delta^{\mu\bar\mu} = -0.03$.
The allowed range takes the $B_s$ mixing constraints into account and falls within the  2\,$\sigma$ C.L of $\overline{R}$ as defined in \eqref{RexpValues}. 
We observe that for $M_H\le 0.75\tev$ negative values of $\mathcal{A}^{\mu\mu}_{\Delta\Gamma}$ are predicted in this scenario, while for $M_H\ge 2.5\tev$ its 
SM value is approached.

\begin{table}[t]
\centering
\begin{tabular}{|c||c|c|c|}
\hline
Model & $M_{H^0} [{\rm GeV}]$ & $S$ & $\overline{R}$ \\
\hline
\hline
MFV & 900 -- 1095 & 1.20 -- 0.81  & 1.22 -- 0.58  \\
LHS, RHS & 1030 -- 1310 & 0.94 -- 0.56 & 0.75 -- 0.26  \\
\hline
\end{tabular}
\caption{Ranges allowed by $B_s$ mixing for the Scalar+Pseudoscalar model with non-degenerate masses. The model has been tuned so that $P=0$, and thereby corresponds to {\it Scenario E}. The pseudoscalar mass is $M_{A^0}=1\,{\rm TeV}$.}
\label{tab:HnAScenE}
\end{table}

We will now consider a mass difference between the scalar and pseudoscalar: 
\begin{equation}
M_{H^0}\neq M_{A^0}.
\end{equation}
We consequently shift our focus from {\it Scenario C} to {\it Scenario E}.
Note, however, that a small mass difference is still consistent with a 2HDM in a decoupling regime,
and will approximately reproduce {\it Scenario C}.
In the presence of a non-zero mass difference the following contribution to the SM box function in MFV or a LHS model no longer vanishes:
\begin{align}
    \left[\Delta S({B_s})\right]_{LL} &= \left(\frac{[\tilde{\Delta}^{sb}]^*}{ V_{ts}V_{tb}^*}\right)^2
    \left[\frac{r^{LL}(M_{H^0})}{M^2_{H^0}} - \frac{r^{LL}(M_{A^0})}{\,M^2_{A^0}} \right]
\end{align}
and an analogous expression for a RHS model after the replacement $[\tilde{\Delta}^{sb}]^*\to \tilde{\Delta}^{bs}$ and $L$ with $\overline{R}$ with $r^{RR}=r^{LL}=50\tev^2$.
In MFV we therefore find:
\begin{align}
    \left[\Delta S({B_s})\right]_{LL} = 
    \left(\frac{m_b}{m_s}\right)^2 \left(\left[\Delta S({B_s})\right]_{RR}\right)^*
    &= (\epsilon^*)^2\, y_b^2\, y_t^4 
    \left[\frac{r^{LL}(M_{H^0})}{M^2_{H^0}} - \frac{r^{LL}(M_{A^0})}{\,M^2_{A^0}} \right]\label{SLL}.
\end{align}

We observe that:
\begin{itemize}
\item
The LL contribution, while suppressed relatively to the LR 
contribution through smaller hadronic matrix element ($|r^{LR}|\approx 6|r^{LL}|$) and the splitting between scalar and pseudoscalar masses, 
does not suffer from $m_s/m_b$ suppression. Thus whether the SLL contribution or 
LR dominates depends sensitively on the size of the mass splitting in question. The SRR contribution is totally negligible.
\item
The SLL contribution now contains in principle a new flavour-blind phase in 
$\epsilon$ allowing for new CP-violating effects in $B_s$ mixing.
\item
The formulae  (\ref{LR}) and (\ref{SLL}) are also valid in a non-MFV framework
in which new flavour and CP-violating phases are present in $\epsilon$.
\end{itemize}

For our numerics we fix the pseudoscalar mass to $M_{A^0}=1\,{\rm TeV}$ and will continue to assume a universal lepton coupling as given in \eqref{HnALeptonCouplings}.
The quark coupling required to set $P=0$ has strength $|\tilde\Delta^{sb}| = 0.0043$ and a phase ${\rm arg}(\tilde\Delta^{sb}) = \phi_s^{\rm SM}/2 = -1^\circ$.
This then poses the question of what the allowed range for the scalar mass $M_{H^0}$ is that is compatible with the $B_s$ mixing constraints, and whether the resulting $S$ satisfies the experimental bounds on $\overline{R}$.

Because the scalar and pseudoscalar couple in the same way to quarks (see \eqref{HnALagEqual}), $P=0$ implies that there are no new CP violating phases in the decay.
The NP mixing phase $\phi_s^{NP}$ can still contribute (unless we assume MFV), and we find for the time-dependent observables:
\begin{equation}
    {\cal A}_{\Delta\Gamma}^{\mu\mu} = -\cos(\phi_s^{\rm NP}),\qquad {\cal S}_{\mu\mu} = \sin(\phi_s^{\rm NP}).
\end{equation}
This change in sign for these observables with respect to the SM is a smoking gun signal of scalars dominating in the $\Bsmumu$ decay.
In Table~\ref{tab:HnAScenE} we summarise our results for the allowed ranges of the scalar mass $M_{H^0}$, parameter $S$ and observable $\overline{R}$ for both MFV and a LHS/RHS quark model.
Both models are seen to satisfy the current experimental range for $\overline{R}$ given in \eqref{RexpValues}.

\section{Summary}\label{Summary}

We have performed the first detailed phenomenological analysis of the time-dependent rate for the $B_s\to\mu^+\mu^-$ decay following the formalism developed in 
Ref.~\cite{deBruyn:2012wk}. 
Our analysis demonstrates that decay-time studies of $B_s\to\mu^+\mu^-$, which offer the observables ${\cal A}^{\mu\mu}_{\Delta\Gamma}$ and ${\cal S}_{\mu\mu}$ in addition to the branching ratio, allow for various NP scenarios to be disentangled.
Specifically, the presence of new scalar, pseudoscalar or gauge boson particles can potentially be identified, which is not possible on the basis of the branching ratio alone.

We have proposed a classification of various NP scenarios in terms of two complex 
variables $S$ and $P$ that fully describe the three observables involved, 
and can be expressed in terms of the fundamental parameters of a given model. 
The experimental determination of $S$ and $P$, 
accompanied by plausible model specific assumptions, will allow us to probe 
NP in this theoretically clean decay. We have illustrated this by placing several popular extensions of the SM into phenomenological scenarios introduced by us (see Table~\ref{tab:Models}).

We have further presented numerical analyses for the observables in question in 
models for tree-level contributions to $B_s\to\mu^+\mu^-$ mediated by 
heavy gauge bosons, scalars and pseudoscalars. 
The plots in Figures~\ref{fig:GRAND1} and~\ref{fig:GRAND2} illustrate our general findings.
Our main messages from these analyses are as follows.
\begin{itemize}
\item 
    The phenomenology of a tree-level $Z'$ exchange with respect to the studied observables is very different in structure to that of spin-0 particles, in particular pseudoscalars.
        This is shown in Figure~\ref{fig:GRAND1} (see also Figure~\ref{fig:Z1} versus Figure~\ref{fig:H2}).
\item
    In turn, the phenomenology of a scalar is more restricted than that of a pseudoscalar. For instance, suppression of $\overline{R}$ with respect to its SM value would exclude a NP scenario with only a single scalar, whereas such a suppression is possible for a single pseudoscalar.
 \item
    For models with a single new particle with a mass of $1\tev$  -- specifically a gauge boson, scalar or pseudoscalar -- negative values of $\mathcal{A}^{\mu\mu}_{\Delta\Gamma}$ require large couplings to muons and a significant deviation of the $B_s$ mixing phase $\phi_s$ from its SM value.
\item
On the contrary, a negative value of $\mathcal{A}^{\mu\mu}_{\Delta\Gamma}$ can naturally be explained in models with both a scalar and pseudoscalar and a common mass $M_H\le 1.5\tev$.
An example is decoupled 2HDMs.
In this setup a deviation of $\phi_s$ from its SM value is not required.
Furthermore, in these models $\overline{R}$ has a strict lower bound.
\item 
For a $Z'$ scenario the required suppression of $\Delta M_s$, due to current tensions with experiment, implies that ${\cal S}_{\mu\mu}\not=0$.
\item
For a single pseudoscalar or scalar scenario, the required suppression of $\Delta M_s$ 
implies a departure from the SM value of ${\rm BR}(\Bsmumu)$.
\end{itemize}

The numerous plots and examples presented by us  provide a roadmap for
future experimental results of this outstanding rare $B$ decay. 

\subsubsection*{Acknowledgements}
We thank Christine Davies, Fulvia De Fazio, Martin Gorbahn, Minoru Nagai, David Straub and Robert Ziegler for discussions.
RK would like to thank Andrzej Buras and his group for hosting him at the IAS/TU in Munich.
This research was financially supported by the ERC Advanced Grant project ``FLAVOUR'' (267104) and the Foundation for Fundamental Research on Matter (FOM).

\bibliographystyle{JHEP}

\bibliography{allrefs}

\end{document}